\documentclass[aps,prl,reprint,twocolumn,superscriptaddress]{revtex4-2}
\usepackage{amsmath,amssymb}
\usepackage{graphicx}
\usepackage{bm}
\usepackage{physics}
\usepackage{color}
\usepackage{tikz}
\usetikzlibrary{decorations.markings}
\usepackage{mathtools}

\NewDocumentCommand{\GM}{o}{%
  \mathrm{GM}\IfValueT{#1}{^{(#1)}}%
}

\begin{document}

\title{Can Multipartite Entanglement Probe Wormhole Moduli\\
Invisible to Bipartite Entanglement?}

\author{Norihiro Iizuka}
\affiliation{Department of Physics, National Tsing Hua University, Hsinchu 300044, Taiwan}
\affiliation{Yukawa Institute for Theoretical Physics, Kyoto University, Kyoto 606-8502, Japan}

\date{\today}

\begin{abstract}
We ask whether multipartite entanglement can probe bulk moduli invisible to bipartite entanglement. We study this question in an explicit family of four-boundary AdS$_3$ wormholes related by a Fenchel--Nielsen twist $\tau$, asking whether the complete set of bipartite RT entropies can remain exactly $\tau$-independent while the holographic $\mathtt q=4$ multi-entropy is sensitive to the twist. Within the class of configurations analyzed here, for sufficiently small $|\tau|$, we find no such regime along the symmetric line $\ell=m$: the globally minimal $\mathtt q=4$ network is itself exactly $\tau$-independent throughout the parameter range analyzed. This conclusion is not a priori obvious: it emerges from a competition among several admissible configurations whose relative ordering changes multiple times as the geometry is varied, and persists up to the point where the bipartite RT entropies themselves become twist-sensitive.
\end{abstract}

\maketitle\section{I. Introduction}

Multipartite entanglement can contain information that is not
captured by any single bipartite cut. In holography, entanglement
entropy is conventionally computed by the Ryu--Takayanagi (RT)
prescription \cite{Ryu:2006bv,Ryu:2006ef}, in which a boundary
region is associated with a bulk minimal surface. By construction,
however, entanglement entropy is bipartite: it probes only the
entanglement between a region and its complement.

This immediately raises a natural question. If two holographic
states, or two bulk geometries, cannot be distinguished by any
bipartite entanglement entropy, is there a more sensitive
entanglement measure that can distinguish them? A natural candidate
is the holographic multi-entropy $S^{(\mathtt q)}$ \cite{Gadde:2022cqi, Penington:2022dhr, Gadde:2023zzj}, and
its genuine multipartite refinement $\mathrm{GM}^{(\mathtt q)}$
\cite{Iizuka:2025ioc,Iizuka:2025caq}, which probe genuinely
multipartite correlations beyond any single bipartite cut.

In this paper we ask this question concretely for a family of
four-boundary AdS$_3$ wormholes related by a Fenchel--Nielsen twist
$\tau$ along an internal pants-decomposition curve: does there exist
a parameter regime in which the complete set of bipartite RT
entropies is exactly $\tau$-independent while the $\mathtt q=4$ multi-entropy
$S^{(4)}$ is $\tau$-dependent? The answer is not obvious a priori:
$S^{(4)}$ is determined by a global minimization over admissible
multiway-cut configurations, and the minimum could in principle be
realized by a $\tau$-independent configuration, in which case
$S^{(4)}$ would fail to detect the twist even though it is not
itself built out of bipartite RT data.

We show that, within the class of admissible configurations analyzed here
(Secs.~IV--VI), for sufficiently small $|\tau|$, the answer is negative
along the symmetric line $\ell=m$:  at every point examined, the globally
minimal $\mathtt q=4$ network is itself exactly $\tau$-independent,
so $S^{(4)}$ fails to detect the twist. At one point (Parameter Choice~1),
this minimal network is built from horizon and cuff geodesics
alone; at a second point (Parameter Choice~2), it is instead a
hybrid configuration in which one pair of pants carries an
optimized Steiner tripod while the other is cut directly along a
horizon. Tracking this competition as $\ell=m$ is varied for sufficiently small
$|\tau|$ (Sec.~VI), we find that whenever the minimal $\tau$-independent configuration might in principle have been overtaken,
either it is not, or the bipartite RT entropies themselves become twist-sensitive first.

\section{II. The Setup}

We consider a four-boundary AdS$_3$ wormhole whose time-symmetric
slice is a hyperbolic surface obtained as a quotient
$\Sigma=\mathbb{H}^2/\Gamma$ by a Fuchsian group
$\Gamma\subset SL(2,\mathbb{R})$.
Such multiboundary wormholes can be constructed by gluing pairs of
pants along internal geodesics \cite{Aminneborg:1997pz, Brill:1998pr, Skenderis:2009ju, Balasubramanian:2014hda}.
For explicit calculations, we use the quotient parametrization used in
Ref.~\cite{Caceres:2019giy, Anegawa:2020lzw}.

\begin{figure*}[t]
\centering

\begin{tikzpicture}[
    x=1.05cm,
    y=1.05cm,
    scale=0.82,
    transform shape,
    line cap=round,
    line join=round,
    arrow along path/.style={
        postaction={
            decorate,
            decoration={
                markings,
                mark=at position 0.55 with
                {\arrow{stealth}}
            }
        }
    }
]

\def\muR{7.389056}   
\def\cOne{5.28094}
\def\rOne{1.3}
\def\cTwo{1.94275}
\def\rTwo{0.9}

\fill[yellow!33]
    (-\muR,0)
    arc[start angle=180,end angle=0,radius=\muR]
    -- cycle;

\fill[white]
    (-1,0)
    arc[start angle=180,end angle=0,radius=1]
    -- cycle;

\fill[white]
    ({\cOne-\rOne},0)
    arc[start angle=180,end angle=0,radius=\rOne]
    -- cycle;

\fill[white]
    ({\cTwo-\rTwo},0)
    arc[start angle=180,end angle=0,radius=\rTwo]
    -- cycle;

\fill[white]
    ({-\cTwo-\rTwo},0)
    arc[start angle=180,end angle=0,radius=\rTwo]
    -- cycle;

\fill[white]
    ({-\cOne-\rOne},0)
    arc[start angle=180,end angle=0,radius=\rOne]
    -- cycle;

\fill[gray!15]
    (2.84275,0)
    -- (3.98094,0)
    arc[start angle=180,end angle=131.2265,radius=1.3]
    arc[start angle=50.2790,end angle=147.8256,radius=1.2711745]
    arc[start angle=48.7734,end angle=0,radius=0.9]
    -- cycle;

\fill[gray!15]
    (-3.98094,0)
    -- (-2.84275,0)
    arc[start angle=180,end angle=131.2266,radius=0.9]
    arc[start angle=32.1744,end angle=129.7210,radius=1.2711745]
    arc[start angle=48.7735,end angle=0,radius=1.3]
    -- cycle;

\fill[gray!15]
    (6.58094,0)
    -- (7.389056,0)
    arc[start angle=0,end angle=16.5902,radius=7.389056]
    arc[start angle=142.3695,end angle=169.5508,radius=3.4554105]
    arc[start angle=28.8205,end angle=0,radius=1.3]
    -- cycle;

\fill[gray!15]
    (1,0)
    -- (1.04275,0)
    arc[start angle=180,end angle=151.1792,radius=0.9]
    arc[start angle=111.9092,end angle=142.3695,radius=0.4676390]
    arc[start angle=16.5902,end angle=0,radius=1]
    -- cycle;

\fill[gray!15]
    (-7.389056,0)
    -- (-6.58094,0)
    arc[start angle=180,end angle=151.1795,radius=1.3]
    arc[start angle=10.4492,end angle=37.6305,radius=3.4554105]
    arc[start angle=163.4098,end angle=180,radius=7.389056]
    -- cycle;

\fill[gray!15]
    (-1.04275,0)
    -- (-1,0)
    arc[start angle=180,end angle=163.4098,radius=1]
    arc[start angle=37.6305,end angle=68.0908,radius=0.4676390]
    arc[start angle=28.8208,end angle=0,radius=0.9]
    -- cycle;

\draw[->,line width=0.8pt]
    (-8.0,0) -- (8.2,0)
    node[right] {$x$};

\draw[->,line width=0.8pt]
    (0,0) -- (0,7.9)
    node[above] {$y$};

\draw[blue!65!black,line width=1.5pt,arrow along path]
    (-\muR,0)
    arc[start angle=180,end angle=0,radius=\muR];

\draw[blue!65!black,line width=1.5pt,arrow along path]
    (-1,0)
    arc[start angle=180,end angle=0,radius=1];

\draw[red!75!black,line width=1.5pt,arrow along path]
    ({\cOne-\rOne},0)
    arc[start angle=180,end angle=0,radius=\rOne];

\draw[red!75!black,line width=1.5pt,arrow along path]
    ({\cTwo+\rTwo},0)
    arc[start angle=0,end angle=180,radius=\rTwo];

\draw[green!50!black,line width=1.5pt,arrow along path]
    ({-\cOne+\rOne},0)
    arc[start angle=0,end angle=180,radius=\rOne];

\draw[green!50!black,line width=1.5pt,arrow along path]
    ({-\cTwo-\rTwo},0)
    arc[start angle=180,end angle=0,radius=\rTwo];

\node[blue!65!black,font=\small] at (0,7.10) {$C_b$};
\node[blue!65!black,font=\small] at (0,0.72) {$C_a$};

\node[red!75!black,font=\small]
    at (\cOne,0.98) {$C_1$};

\node[red!75!black,font=\small]
    at (\cTwo,0.62) {$C_2$};

\node[green!50!black,font=\small]
    at (-\cTwo,0.62) {$C_2'$};

\node[green!50!black,font=\small]
    at (-\cOne,0.98) {$C_1'$};

\node[blue!65!black,font=\small]
    at (0,6.55)
    {$C_a \xleftrightarrow{\ \gamma_1\ } C_b$};

\node[red!75!black,font=\small]
    at (4.15,1.80)
    {$C_1 \xleftrightarrow{\ \gamma_2\ } C_2$};

\node[green!50!black,font=\small]
    at (-4.15,1.80)
    {$C_1' \xleftrightarrow{\ \gamma_2'\ } C_2'$};

\draw[green!50!black,line width=3.2pt]
    (-7.389056,0) -- (-6.58094,0);

\draw[green!50!black,line width=3.2pt]
    (-1.04275,0) -- (-1,0);

\draw[orange,line width=3.2pt]
    (-3.98094,0) -- (-2.84275,0);

\draw[blue,line width=3.2pt]
    (1,0) -- (1.04275,0);

\draw[blue,line width=3.2pt]
    (6.58094,0) -- (7.389056,0);

\draw[red,line width=3.2pt]
    (2.84275,0) -- (3.98094,0);

\node[green!50!black,font=\small,below=8pt]
    at ({(-7.389056-6.58094)/2},0) {$B'$};

\node[green!50!black,font=\small,below=8pt]
    at ({(-1.04275-1)/2},0) {$B'$};

\node[orange,font=\small,below=8pt]
    at ({(-3.98094-2.84275)/2},0) {$B$};

\node[blue,font=\small,below=8pt]
    at ({(1+1.04275)/2},0) {$A'$};

\node[blue,font=\small,below=8pt]
    at ({(6.58094+7.389056)/2},0) {$A'$};

\node[red,font=\small,below=8pt]
    at ({(2.84275+3.98094)/2},0) {$A$};

\foreach \x in {
    -7.389056,-6.58094,-3.98094,-2.84275,
    -1.04275,-1,
     1,1.04275,2.84275,3.98094,6.58094,7.389056
}{
    \fill (\x,0) circle (1.15pt);
}

\node[below=20pt,font=\scriptsize]
    at (-7.389056,0) {$-e^2$};

\node[below=20pt,font=\scriptsize]
    at (-6.58094,0) {$-6.581$};

\node[below=20pt,font=\scriptsize]
    at (-3.98094,0) {$-3.981$};

\node[below=20pt,font=\scriptsize]
    at (-2.84275,0) {$-2.843$};

\node[below=20pt,font=\scriptsize]
    at (-1.04275,0) {$-1.043$};

\node[below=20pt,font=\scriptsize]
    at (-1,0) {$-1$};

\node[below=20pt,font=\scriptsize]
    at (1,0) {$1$};

\node[below=20pt,font=\scriptsize]
    at (1.04275,0) {$1.043$};

\node[below=20pt,font=\scriptsize]
    at (2.84275,0) {$2.843$};

\node[below=20pt,font=\scriptsize]
    at (3.98094,0) {$3.981$};

\node[below=20pt,font=\scriptsize]
    at (6.58094,0) {$6.581$};

\node[below=20pt,font=\scriptsize]
    at (7.389056,0) {$e^2$};

\draw[black,line width=1.6pt]
    (2.535885,0.676898)
    arc[
        start angle=147.8256,
        end angle=50.2790,
        radius=1.2711745
    ];

\node[black,font=\small]
    at (3.56,1.43) {$H_A$};

\draw[black,line width=1.6pt]
    (6.419914,0.626687)
    arc[
        start angle=169.5508,
        end angle=142.3695,
        radius=3.4554105
    ];

\node[black,font=\small]
    at (6.73,2.05) {$H_{A'}$};

\draw[black,line width=1.6pt]
    (-4.424191,0.977743)
    arc[
        start angle=129.7210,
        end angle=32.1744,
        radius=1.2711745
    ];

\node[black,font=\small]
    at (-3.56,1.43) {$H_B$};

\draw[black,line width=1.6pt]
    (-7.081459,2.109761)
    arc[
        start angle=37.6305,
        end angle=10.4492,
        radius=3.4554105
    ];

\node[black,font=\small]
    at (-6.73,1.95) {$H_{B'}$};

\draw[black,line width=1.6pt]
    (1.154231,0.433864)
    arc[
        start angle=111.9092,
        end angle=142.3695,
        radius=0.4676390
    ];

\draw[black,line width=1.6pt]
    (-0.958371,0.285525)
    arc[
        start angle=37.6305,
        end angle=68.0908,
        radius=0.4676390
    ];

\end{tikzpicture}

\caption{
Schottky fundamental region at $\tau=0$.
The colored intervals on the real axis denote the four asymptotic
boundaries $A,A',B,B'$, while the black geodesic arcs
$H_A,H_{A'},H_B,H_{B'}$ are the corresponding horizons.
The gray regions are the four asymptotic funnels.
This figure illustrates Parameter Choice~1
($\ell=m=2$; Sec.~III), for which the circle radii are of
comparable size and the geometry is particularly convenient to
draw; Parameter Choice~2 has the same qualitative structure, but a
to-scale drawing would strongly compress the central region ($C_2$
would be barely visible next to $C_1$ and $C_b$). (Figure prepared
with the assistance of Claude.)
}
\label{fig:schottky-fundamental-region}

\end{figure*}
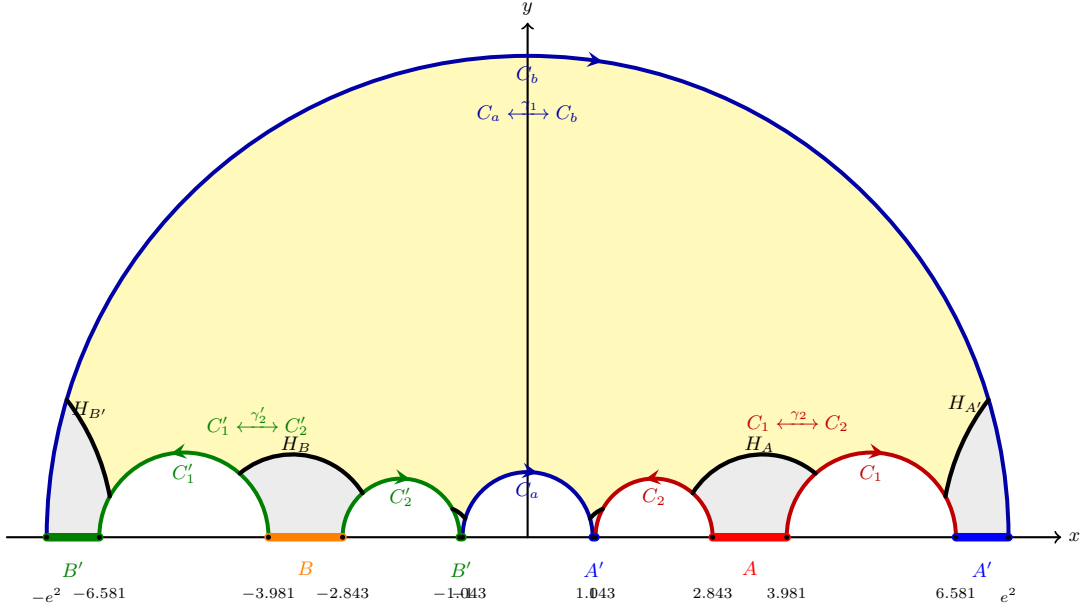

We take
$\Gamma=\langle\gamma_1,\gamma_2,\gamma_2'\rangle$.
A fundamental domain is bounded by six disjoint geodesic
semicircles centered on the real axis,
\begin{align}
C_a: |z|=1,  \qquad \quad \,\,\,\,\, & C_b: |z|=\mu^2,\\
C_1: |z-c_1|=R_1, \quad & C_2: |z-c_2|=R_2,
\end{align}
together with $C_1',C_2'$, defined analogously by
$c_1',c_2',R_1',R_2'$, as shown in
Fig.~\ref{fig:schottky-fundamental-region}.
The generator $\gamma_1$ identifies $C_a\to C_b$ by a pure
dilatation and preserves the orientation along the paired circles,
whereas $\gamma_2$ and $\gamma_2'$ identify
$C_1\to C_2$ and $C_1'\to C_2'$, respectively, with reversed
orientation along the paired circles. Explicitly,
\begin{align}
\gamma_1 &=
\begin{pmatrix}
\mu & 0\\
0 & \mu^{-1}
\end{pmatrix},
\\
\label{gamma2matrixrepn}
\gamma_2 &=
\frac{1}{\sqrt{R_1R_2}}
\begin{pmatrix}
-c_2 & c_1c_2+R_1R_2\\
-1 & c_1
\end{pmatrix},
\\
\gamma_2' &=
\frac{1}{\sqrt{R_1'R_2'}}
\begin{pmatrix}
-c_2' & c_1'c_2'+R_1'R_2'\\
-1 & c_1'
\end{pmatrix}.
\end{align}
The corresponding identifications are illustrated in
Fig.~\ref{fig:schottky-fundamental-region} and verified explicitly
in Appendix~A.
We impose the $\mathbb{Z}_2$-symmetric choice
\begin{equation}
c_1'=-c_1,\quad
c_2'=-c_2,\quad
R_1'=R_1,\quad
R_2'=R_2 .
\end{equation}

For a hyperbolic element $\gamma\in SL(2,\mathbb{R})$, the length of
the corresponding closed geodesic is determined by 
\begin{equation}
2\cosh\frac{L(\gamma)}{2}=|\Tr \gamma| .
\label{tracelengthformula}
\end{equation}
See Appendix B. 
The four asymptotic horizons are therefore
\begin{align}
\label{eq:LAequalell}
L_A
 &=L(\gamma_2)
 =2\cosh^{-1}\!\left|
 \frac{c_1-c_2}{2\sqrt{R_1R_2}}
 \right|,
\\
L_{A'}
\label{eq:LAprimeequalell}
 &=L(\gamma_1\gamma_2)
 =2\cosh^{-1}\! 
 \left|
 \frac{c_1/\mu-c_2\mu}{2\sqrt{R_1R_2}}
 \right|
 ,
\\
L_B
 &=L(\gamma_2')
 =2\cosh^{-1}\!\left|
 \frac{c_1'-c_2'}{2\sqrt{R_1'R_2'}}
 \right|,
\\
L_{B'}
 &=L(\gamma_1\gamma_2')
 =2\cosh^{-1}\!
 \left|
 \frac{c_1'/\mu-c_2'\mu}{2\sqrt{R_1'R_2'}}
 \right|
.
\end{align}
The internal geodesic along which the two pairs of pants are glued has
length
\begin{equation}
\label{eq:LMequalm}
L_M=L(\gamma_1)=2\log\mu .
\end{equation}
With the above $\mathbb{Z}_2$ symmetry,
\begin{equation}
L_A=L_B,\qquad L_{A'}=L_{B'} .
\end{equation}

The Fenchel--Nielsen twist is implemented by conjugation~\cite{Wolpert:1982, Goldman:2004csg},
\begin{equation}
\gamma_2'(\tau)
=
\eta(\tau)\gamma_2'\eta(-\tau),
\quad
\eta(\tau)
=
\begin{pmatrix}
e^{\tau/2}&0\\
0&e^{-\tau/2}
\end{pmatrix}.
\end{equation}
Here $\eta(\tau)$ is the hyperbolic translation by signed length
$\tau$ along the axis of $\gamma_1$. Cutting $\Sigma$ along $M$ and
conjugating the holonomies of the pair of pants containing $B,B'$ by
$\eta(\tau)$, while leaving the pair of pants containing $A,A'$
fixed, shifts their relative gluing along $M$ without disturbing the
cuff itself. Since $\eta(\tau)\gamma_1\eta(\tau)^{-1}=\gamma_1$, only
the independent generator $\gamma_2'$ need be rewritten.

Since the length of a closed geodesic associated with a hyperbolic
element $\gamma$ is determined only by $|\Tr\gamma|$, conjugation leaves
its length invariant. Therefore
\begin{equation}
L_B(\tau)
=
L\!\left(\gamma_2'(\tau)\right)
=
L(\gamma_2')
=
L_B .
\end{equation}
Moreover, because $\eta(\tau)$ commutes with $\gamma_1$,
\begin{equation}
\gamma_1\gamma_2'(\tau)
=\gamma_1\eta(\tau)\gamma_2'\eta(-\tau)
=
\eta(\tau)\gamma_1\gamma_2'\eta(-\tau),
\end{equation}
and hence
\begin{equation}
L_{B'}(\tau)
=
L\!\left(\gamma_1\gamma_2'(\tau)\right)
=
L(\gamma_1\gamma_2')
=
L_{B'} .
\end{equation}
The generators $\gamma_1$ and $\gamma_2$ are not changed by the twist,
so $L_M$, $L_A$, and $L_{A'}$ are trivially independent of $\tau$.
Thus
\begin{equation}
L_A,\qquad L_{A'},\qquad L_B,\qquad L_{B'},\qquad L_M
\end{equation}
are all exactly independent of the Fenchel--Nielsen twist parameter
$\tau$. This invariance provides a direct consistency check that
$\eta(\tau)$ implements a twist rather than some other deformation.

In the following, we consider the symmetric family
\begin{equation}
L_A=L_{A'}=L_B=L_{B'}=\ell,\qquad L_M=m,
\end{equation}
with $\ell$ and $m$ fixed, and vary only the Fenchel--Nielsen modulus
$\tau$, in units of the AdS radius, $L_{\rm AdS}=1$. Newton's
constant $G_N$ is kept explicit throughout, entering only via the
entropy formulas $S=\mathrm{Length}/4G_N$.

\begin{figure}[t]
\centering
\includegraphics[width=0.49\textwidth]{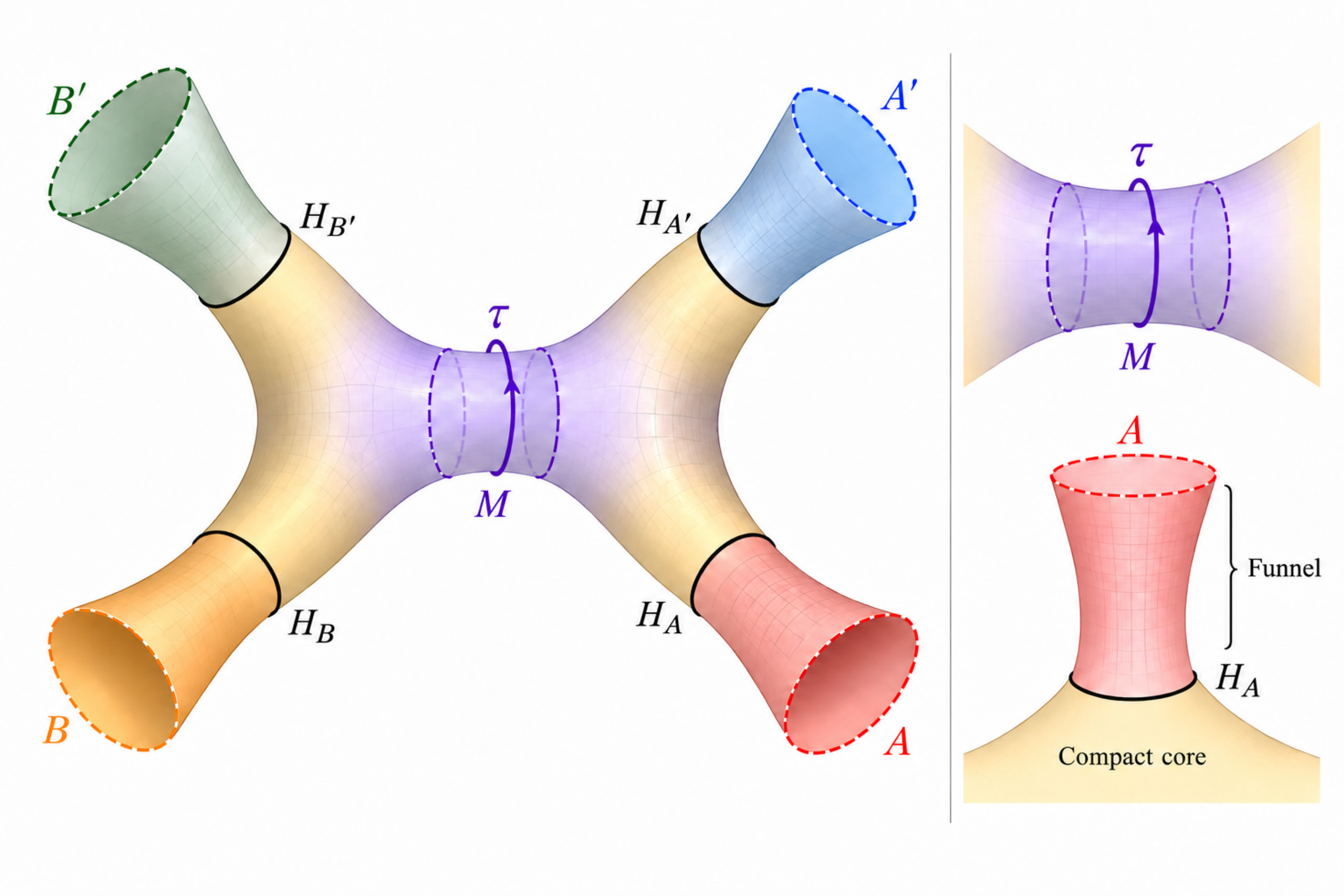}
\caption{
Schematic geometry of the four-boundary wormhole.
The horizons $H_A,H_{A'},H_B,H_{B'}$ bound the compact core,
with funnels extending to the asymptotic boundaries
$A,A',B,B'$.
The two pairs of pants are glued along the internal cuff $M$
with Fenchel--Nielsen twist $\tau$. (Figure prepared with the assistance of ChatGPT.)
}
\label{fig:four-boundary-wormhole}
\end{figure}

\section{III. Why Bipartite Entanglement Is Not Enough}

Although the Fenchel--Nielsen twist changes the hyperbolic geometry,
the complete set of bipartite RT entropies remains unchanged in the
parameter range considered here. The reason is that the geodesics
that are sensitive to the twist never dominate the corresponding RT
minimization.

For the four single-boundary regions, the RT surfaces are simply the
four horizons. Hence
\begin{equation}
S(A)=S(A')=S(B)=S(B')=\frac{\ell}{4G_N},
\end{equation}
which are exactly independent of $\tau$.

It remains to consider the two-boundary regions. Since the total
four-boundary state is pure, it is sufficient to consider the three
inequivalent choices $AA'$, $AB$, and $AB'$.

For $AA'$, the two competing RT configurations are the two individual
horizons $A\cup A'$ and the internal geodesic $M$. Therefore
\begin{equation}
S(AA')
=
\frac{1}{4G_N}
\min\{2\ell,m\},
\end{equation}
independent of $\tau$ for either value of the minimum, since a
Fenchel--Nielsen twist changes only the relative gluing of the two
pairs of pants along $M$, leaving the lengths of the individual
horizons and of $M$ unchanged (Sec.~II).

The remaining cases, $AB$ and $AB'$, are the only ones for which a
connected candidate can probe the relative twist of the two pairs of pants. We now derive a condition under which no such connected
candidate can dominate the RT minimization.

Recall that $M$ separates $\{A,A'\}$ from $\{B,B'\}$. Let $\Gamma$ be a simple closed curve realizing either crossing
partition $AB|A'B'$ or $AB'|A'B$. Such a curve cannot be disjoint
from $M$: if it were, it would lie entirely within one of the two
pairs of pants, where every essential simple closed curve is
boundary-parallel, and hence could not realize a crossing partition.
Let $i(\Gamma,M)$ denote the geometric intersection number between
the two simple closed curves $\Gamma$ and $M$. Then
$i(\Gamma,M)>0$.
Since both $\Gamma$ and $M$ are separating curves, their mod-$2$
intersection number vanishes, so their geometric intersection number
is even. Therefore
\begin{equation}
i(\Gamma,M)\geq2 .
\end{equation}

Cutting along $M$, the intersection of $\Gamma$ with each pair of
pants must contain at least one essential returning arc from $M$ to
itself separating the two external cuffs on that side. The shortest
representative in the corresponding proper homotopy class is the
unique orthogeodesic perpendicular to $M$ at both endpoints. Denoting
its lengths on the two sides by $z_R$ and $z_L$, we obtain
\begin{equation}
L(\Gamma)\geq z_R+z_L .
\end{equation}
Since the Fenchel--Nielsen twist changes only the gluing along $M$
and leaves the intrinsic geometry of each pair of pants unchanged,
the returning orthogeodesic lengths are independent of $\tau$.
Moreover, the two pairs of pants are isometric in the symmetric family,
so that
\begin{equation}
z_R=z_L\equiv z .
\end{equation}
and hence
\begin{equation}
L(\Gamma)\geq2z .
\end{equation}

The returning orthogeodesic can be evaluated directly for any $\ell,m$.

Choose the lift (to the universal cover $\mathbb H^2$) of the internal cuff $M$ to be the imaginary axis,
\begin{equation}
\widetilde M=\{x=0\},
\end{equation}
whose ideal endpoints are $0$ and $\infty\,(=i\infty)$.

The returning orthogeodesic in the pair of pants bounded by $A$,
$A'$, and $M$ separates $A$ from $A'$; equivalently, closing it with
a segment of $M$ gives a loop in the homotopy class of the
$A$-cycle. Lifting this closed loop to $\mathbb H^2$, going once around the
$A$-cycle acts by the deck transformation $\gamma_2$. Hence the
lifted returning arc joins $\widetilde M$ to
$\gamma_2(\widetilde M)$. 

Therefore the shortest returning arc is the common perpendicular
between these two geodesics,
\begin{equation}
z_R
=
d\!\left(\widetilde M,\gamma_2(\widetilde M)\right).
\end{equation}
Equivalently, one may close the returning arc on the other side of
$M$, obtaining the $A'$-cycle, represented by
$\gamma_1\gamma_2$. Since $\gamma_1$ preserves $\widetilde M$,
\begin{equation}
d\!\left(\widetilde M,
(\gamma_1\gamma_2)(\widetilde M)\right)
=
d\!\left(\widetilde M,
\gamma_2(\widetilde M)\right)
=
z_R .
\end{equation}
Thus the two choices give the same returning-orthogeodesic length, and we use the $A$-cycle representative throughout without loss of generality.
It is important that this is a distance between two geodesics:
the endpoints of the lifted returning arc are not fixed points related
by $\gamma_2$. In particular, $z_R$ is not the translation length
$L(\gamma_2)$.

The ideal endpoints of $\gamma_2(\widetilde M)$ are obtained by acting
with $\gamma_2$ on $0$ and $\infty\,(=i\infty)$. Using Eq.~\eqref{gamma2matrixrepn} 
(or Eq.~\eqref{gamma2zformula}), 
we find
\begin{equation}
\gamma_2( \infty)=c_2,
\qquad
\gamma_2(0)=c_2+\frac{R_1R_2}{c_1}.
\end{equation}
Thus, writing
\begin{equation}
r=c_2,
\qquad
s=c_2+\frac{R_1R_2}{c_1},
\end{equation}
the two geodesics are $(0,\infty)$ and $(r,s)$, and their common
perpendicular has length
\begin{equation}
\begin{aligned}
z_R
&=
\operatorname{arcosh}\!\left(\frac{r+s}{s-r}\right)
=
\log\!\left(
\frac{\sqrt{s}+\sqrt{r}}
{\sqrt{s}-\sqrt{r}}
\right) \\
&=\operatorname{arcosh}\!\left(
1+\frac{2c_1c_2}{R_1R_2}
\right)
=
\operatorname{arcosh}\!\left[
1+\frac{2\cosh^2(\ell/2)}{\sinh^2(m/4)}
\right],
\label{eq:zformula}
\end{aligned}
\end{equation}
where in the last equality we used Eq.~\eqref{eq:LAequalell}, \eqref{eq:LAprimeequalell} and \eqref{eq:LMequalm}. 
See Appendix C and footnote~[30] 
as well. 
This expresses $z=z(\ell,m)$ as an explicit function of the pants
geometry alone.
Let $L^{\rm conn}_{AB}(\tau)$ denote the length of the shortest
\emph{connected} curve realizing the partition $AB|A'B'$---as
opposed to the disconnected candidate $A\cup B$ itself---and define
$L^{\rm conn}_{AB'}(\tau)$ analogously for $AB'|A'B$. Since the bound
$L(\Gamma)\geq z_R+z_L=2z$ holds for every simple closed curve
$\Gamma$ realizing either crossing partition, independently of its homotopy class, it holds in particular for the
length-minimizing representative in each class. Hence
\begin{equation}
L^{\rm conn}_{AB}(\tau)\geq2z,\qquad
L^{\rm conn}_{AB'}(\tau)\geq2z,
\end{equation}
independently of $\tau$. Consequently, whenever
\begin{equation}
2z(\ell,m)>2\ell,
\label{eq:crossing-condition}
\end{equation}
no connected candidate can dominate the RT minimization, so
\begin{equation}
S(AB)=S(AB')=\frac{2\ell}{4G_N}.
\end{equation}
The complete independent set of bipartite entropies is then
\begin{align}
\label{eq:SAetc}
& S(A)= S(A')= S(B)= S(B')  = \frac{\ell}{4G_N},
\\
& S(AA')= \frac{\min\{2\ell,m\}}{4G_N},
\quad  S(AB)= S(AB')= \frac{2\ell}{4G_N},
\label{eq:SABetc}
\end{align}
with all remaining two-boundary entropies fixed by purity, and none
of these quantities depends on the Fenchel--Nielsen twist $\tau$.
The complete bipartite entropy vector is therefore exactly constant
along this one-parameter family,
\begin{equation}
\partial_\tau \mathbf{S}_{\rm RT}=0,
\end{equation}
even though the underlying hyperbolic geometry varies with $\tau$.
The twist modulus is therefore invisible to all bipartite RT
entropies whenever Eq.~\eqref{eq:crossing-condition} holds.

In particular, since $\partial_\tau\mathbf S_{\rm RT}=0$, so is any
linear combination of these bipartite entropies. This includes the
tripartite information
\begin{equation}
\begin{aligned}
I_3(A{:}A'{:}B)&=S(A)+S(A')+S(B)+S(B') \\
&\qquad -S(AA')-S(AB)-S(A'B),
\end{aligned}
\end{equation}
which is a diagnostic for the $\mathtt q=4$-partite case \cite{Balasubramanian:2014hda,Iizuka:2025caq}.
Using Eqs.~\eqref{eq:SAetc} and \eqref{eq:SABetc}, we find
$I_3(A{:}A'{:}B)=-\min\{2\ell,m\}/(4G_N)$, exactly independent of $\tau$ and
consistent with the general holographic bound
$I_3\leq0$~\cite{Hayden:2011ag}.  Detecting the twist
therefore requires an entanglement measure not reducible to any
combination of ordinary bipartite entropies \footnote{Related multipartite diagnostics constructed from ordinary
RT entropies or from RT surfaces in canonically purified geometries,
including $I_3$, $R_3$, and $Q_4$, have been systematically studied
for multiboundary wormholes in
Ref.~\cite{Balasubramanian:2024ysu}. These are distinct from the holographic multi-entropy
$S^{(4)}$ studied here.}.

\paragraph*{Two parameter choices.}
To make the comparison concrete, we now restrict further to the
one-dimensional slice $\ell=m$ within the symmetric family of
Sec.~II, and work throughout with two representative points on this
line, chosen so that Eq.~\eqref{eq:crossing-condition} holds at
both, so that their complete bipartite RT entropy vectors are
exactly $\tau$-independent. Which configuration minimizes the $\mathtt
q=4$ multi-entropy at each point is a separate question, addressed
in Sec.~V.

\emph{Parameter Choice 1:}
\begin{equation}
\begin{aligned}
&\ell=m=2,\quad
\mu=e,\quad R_1=1.3,\quad R_2=0.9, \\
&c_1=5.280945\ldots,\qquad c_2=1.942751\ldots,
\end{aligned}
\end{equation}
for which $2z=7.224452\ldots>2\ell=4$ (from Eq.~\eqref{eq:zformula}), so 
Eq.~\eqref{eq:crossing-condition} holds and
\begin{equation}
\begin{aligned}
& S(A)= S(A')= S(B)= S(B') =\frac{2}{4G_N},\\
& S(AA')=\frac{2}{4G_N}, \quad S(AB)=S(AB')=\frac{4}{4G_N},
\end{aligned}
\end{equation}
with $I_3(A{:}A'{:}B)=-2/(4G_N)$.

$\vspace{1mm}$

\emph{Parameter Choice 2:}
\begin{equation}
\begin{aligned}
& \ell =m=3.55,\quad
\mu=e^{1.775}, \quad R_1 =5.85,\quad R_2=0.20, \\
&c_1=7.905266\ldots,\qquad c_2=1.339812\ldots,
\end{aligned}
\end{equation}
for which $2z=7.284849\ldots>2\ell=7.10$ (from Eq.~\eqref{eq:zformula}), so
Eq.~\eqref{eq:crossing-condition} holds and
\begin{equation}
\begin{aligned}
& S(A)= S(A')= S(B)= S(B')=\frac{3.55}{4G_N},\\
& S(AA')=\frac{3.55}{4G_N},\quad S(AB)=S(AB')=\frac{7.10}{4G_N},
\end{aligned}
\end{equation}
with $I_3(A{:}A'{:}B)=-3.55/(4G_N)$.

Both parameter choices satisfy
\begin{equation}
\label{parameterconsistency}
1<c_2-R_2<c_2+R_2<c_1-R_1<c_1+R_1<\mu^2,
\end{equation}
so that the six Schottky circles are mutually disjoint and ordered
as in Fig.~\ref{fig:schottky-fundamental-region}. Thus the standard
disjoint-circle construction applies, giving the discrete Schottky
group $\Gamma$ used above.

For both choices, the complete bipartite RT entropy vector is
therefore exactly $\tau$-independent. Figure~\ref{fig:schottky-fundamental-region} illustrates Parameter
Choice~1, for which the geometry is particularly convenient to
visualize.

\section{IV. The Seam-Anchored $\mathtt q=4$ Network}

The holographic multi-entropy $S^{(\mathtt q)}$ is given by a
global minimization over admissible multiway-cut configurations in
the bulk \cite{Gadde:2022cqi}. Among the connected configurations
relevant for this minimization are Steiner networks composed of
geodesic segments meeting at equiangular ($120^\circ$) trivalent
junctions. For $\mathtt q=4$, the two non-crossing connected
topologies, the $s$- and $t$-channels of Fig.~17 in Ref.~\cite{Gadde:2023zzj}, are H-trees, each consisting of two trivalent
vertices joined by a single internal edge. Such Steiner-network
configurations have been extensively used and studied in recent
work; see, for example,
\cite{Harper:2024ker,Gadde:2024taa,Iizuka:2025elr,Iizuka:2025caq,Balasubramanian:2025hxg,Akella:2025owv,Iizuka:2026ahd,Balasubramanian:2026chr,Akella:2026bci,Hu:2026bhg}.

We now construct the H-tree network explicitly for our family and
show how it must be assembled into a global network on the compact
core.

\paragraph*{Horizons versus asymptotic boundaries.}
The four horizons $H_A$, $H_{A'}$, $H_B$, $H_{B'}$ are finite-length closed
geodesics in the bulk, associated with the hyperbolic elements
$\gamma_2$, $\gamma_1\gamma_2$, $\gamma_2'(\tau)$, and
$\gamma_1\gamma_2'(\tau)$ respectively. Each is connected to its
corresponding asymptotic boundary component by an infinite hyperbolic
funnel (see Fig.~\ref{fig:four-boundary-wormhole}). The multi-entropy network is obtained by minimizing over
configurations that lie on the compact core of $\Sigma$ and correctly
realize the required four-region domain partition; extending a
candidate curve out through a funnel to the asymptotic boundary only
adds length and is never minimal, so we restrict attention throughout
to the compact core.

Note that within the compact core, admissible
configurations are not limited to the connected, trivalent H-trees
constructed below: as we show in Sec.~V, \emph{branchless}
configurations (unions of disjoint simple closed geodesics, which
may include the horizons themselves), as well as \emph{hybrid} configurations combining trivalent and branchless components, also realize valid
domain partitions and must be compared on equal footing with the
H-trees in the global minimization defining $S^{(4)}$.

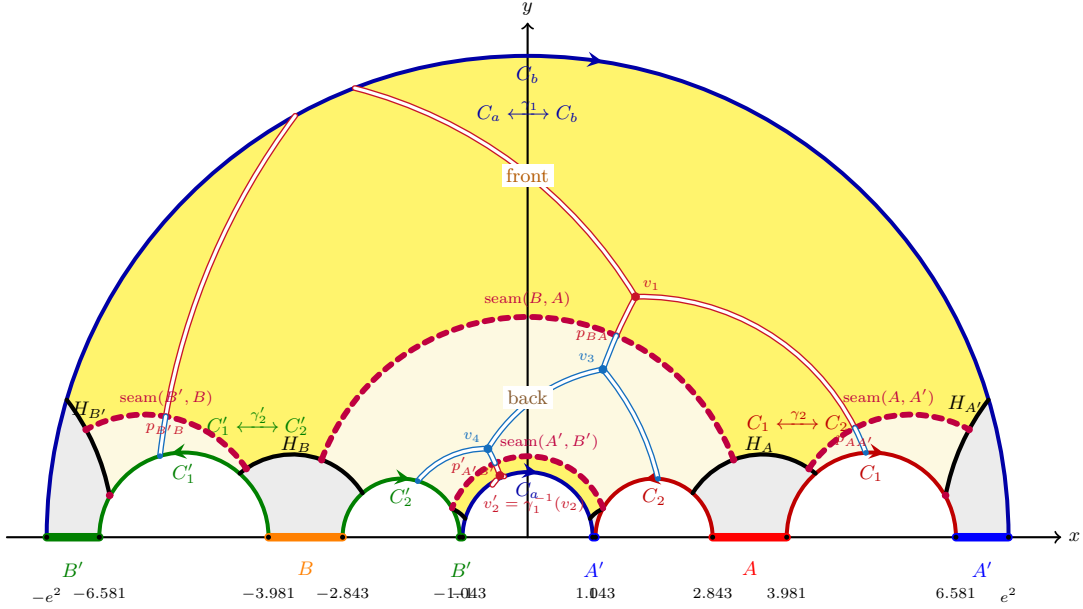
\begin{figure*}[t]
\centering

\begin{tikzpicture}[
    x=1.05cm,
    y=1.05cm,
    scale=0.82,
    transform shape,
    line cap=round,
    line join=round,
    arrow along path/.style={
        postaction={
            decorate,
            decoration={
                markings,
                mark=at position 0.55 with
                {\arrow{stealth}}
            }
        }
    }
]

\def\muR{7.389056}
\def\cOne{5.28094}
\def\rOne{1.3}
\def\cTwo{1.94275}
\def\rTwo{0.9}

\definecolor{htreecolor}{RGB}{200,20,40}
\definecolor{backcolor}{RGB}{20,110,190}

\fill[yellow!70]
    (-\muR,0)
    arc[start angle=180,end angle=0,radius=\muR]
    -- cycle;

\fill[white]
    (-1,0)
    arc[start angle=180,end angle=0,radius=1]
    -- cycle;

\fill[white]
    ({\cOne-\rOne},0)
    arc[start angle=180,end angle=0,radius=\rOne]
    -- cycle;

\fill[white]
    ({\cTwo-\rTwo},0)
    arc[start angle=180,end angle=0,radius=\rTwo]
    -- cycle;

\fill[white]
    ({-\cTwo-\rTwo},0)
    arc[start angle=180,end angle=0,radius=\rTwo]
    -- cycle;

\fill[white]
    ({-\cOne-\rOne},0)
    arc[start angle=180,end angle=0,radius=\rOne]
    -- cycle;

\fill[gray!15]
    (2.84275,0) -- (3.98094,0)
    arc[start angle=180,end angle=131.2265,radius=1.3]
    arc[start angle=50.2790,end angle=147.8256,radius=1.2711745]
    arc[start angle=48.7734,end angle=0,radius=0.9]
    -- cycle;

\fill[gray!15]
    (-3.98094,0) -- (-2.84275,0)
    arc[start angle=180,end angle=131.2266,radius=0.9]
    arc[start angle=32.1744,end angle=129.7210,radius=1.2711745]
    arc[start angle=48.7735,end angle=0,radius=1.3]
    -- cycle;

\fill[gray!15]
    (6.58094,0) -- (7.389056,0)
    arc[start angle=0,end angle=16.5902,radius=7.389056]
    arc[start angle=142.3695,end angle=169.5508,radius=3.4554105]
    arc[start angle=28.8205,end angle=0,radius=1.3]
    -- cycle;

\fill[gray!15]
    (1,0) -- (1.04275,0)
    arc[start angle=180,end angle=151.1792,radius=0.9]
    arc[start angle=111.9092,end angle=142.3695,radius=0.4676390]
    arc[start angle=16.5902,end angle=0,radius=1]
    -- cycle;

\fill[gray!15]
    (-7.389056,0) -- (-6.58094,0)
    arc[start angle=180,end angle=151.1795,radius=1.3]
    arc[start angle=10.4492,end angle=37.6305,radius=3.4554105]
    arc[start angle=163.4098,end angle=180,radius=7.389056]
    -- cycle;

\fill[gray!15]
    (-1.04275,0) -- (-1,0)
    arc[start angle=180,end angle=163.4098,radius=1]
    arc[start angle=37.6305,end angle=68.0908,radius=0.4676390]
    arc[start angle=28.8208,end angle=0,radius=0.9]
    -- cycle;

\fill[yellow!75!brown!15]
    (2.535885,0.676898)
    arc[start angle=147.8256,end angle=110.6065,radius=1.2711745]
    arc[start angle=20.606,end angle=159.394,radius=3.38076]
    arc[start angle=69.3935,end angle=32.1744,radius=1.2711745]
    arc[start angle=131.2266,end angle=29.7035,radius=0.9]
    arc[start angle=158.9875,end angle=21.0125,radius=1.243713]
    arc[start angle=150.2965,end angle=48.7734,radius=0.9]
    -- cycle;

\fill[yellow!75!brown!15]
    (4.424191,0.977743)
    arc[start angle=50.2790,end angle=55.9682,radius=1.2711745]
    arc[start angle=145.968,end angle=61.421,radius=1.88231]
    arc[start angle=151.4211,end angle=169.2941,radius=3.4554105]
    arc[start angle=79.2942,end angle=80.4196,radius=0.653277]
    arc[start angle=29.7035,end angle=131.2265,radius=1.3]
    -- cycle;

\fill[yellow!75!brown!15]
    (-4.424191,0.977743)
    arc[start angle=129.7210,end angle=124.0318,radius=1.2711745]
    arc[start angle=34.032,end angle=118.579,radius=1.88231]
    arc[start angle=28.5789,end angle=10.7059,radius=3.4554105]
    arc[start angle=100.7058,end angle=99.5804,radius=0.653277]
    arc[start angle=150.2965,end angle=48.7735,radius=1.3]
    -- cycle;

\draw[->,line width=0.8pt]
    (-8.0,0) -- (8.2,0)
    node[right] {$x$};

\draw[->,line width=0.8pt]
    (0,0) -- (0,7.9)
    node[above] {$y$};

\draw[
    blue!65!black,
    line width=1.5pt,
    arrow along path
]
    (-\muR,0)
    arc[start angle=180,end angle=0,radius=\muR];

\draw[
    blue!65!black,
    line width=1.5pt,
    arrow along path
]
    (-1,0)
    arc[start angle=180,end angle=0,radius=1];

\draw[
    red!75!black,
    line width=1.5pt,
    arrow along path
]
    ({\cOne-\rOne},0)
    arc[start angle=180,end angle=0,radius=\rOne];

\draw[
    red!75!black,
    line width=1.5pt,
    arrow along path
]
    ({\cTwo+\rTwo},0)
    arc[start angle=0,end angle=180,radius=\rTwo];

\draw[
    green!50!black,
    line width=1.5pt,
    arrow along path
]
    ({-\cOne+\rOne},0)
    arc[start angle=0,end angle=180,radius=\rOne];

\draw[
    green!50!black,
    line width=1.5pt,
    arrow along path
]
    ({-\cTwo-\rTwo},0)
    arc[start angle=180,end angle=0,radius=\rTwo];

\node[blue!65!black,font=\small]
    at (0,7.10) {$C_b$};

\node[blue!65!black,font=\small]
    at (0,0.72) {$C_a$};

\node[red!75!black,font=\small]
    at (\cOne,0.98) {$C_1$};

\node[red!75!black,font=\small]
    at (\cTwo,0.62) {$C_2$};

\node[green!50!black,font=\small]
    at (-\cTwo,0.62) {$C_2'$};

\node[green!50!black,font=\small]
    at (-\cOne,0.98) {$C_1'$};

\node[blue!65!black,font=\small]
    at (0,6.55)
    {$C_a \xleftrightarrow{\ \gamma_1\ } C_b$};

\node[red!75!black,font=\small]
    at (4.15,1.80)
    {$C_1 \xleftrightarrow{\ \gamma_2\ } C_2$};

\node[green!50!black,font=\small]
    at (-4.15,1.80)
    {$C_1' \xleftrightarrow{\ \gamma_2'\ } C_2'$};

\draw[green!50!black,line width=3.2pt]
    (-7.389056,0) -- (-6.58094,0);

\draw[green!50!black,line width=3.2pt]
    (-1.04275,0) -- (-1,0);

\draw[orange,line width=3.2pt]
    (-3.98094,0) -- (-2.84275,0);

\draw[blue,line width=3.2pt]
    (1,0) -- (1.04275,0);

\draw[blue,line width=3.2pt]
    (6.58094,0) -- (7.389056,0);

\draw[red,line width=3.2pt]
    (2.84275,0) -- (3.98094,0);

\node[green!50!black,font=\small,below=8pt]
    at ({(-7.389056-6.58094)/2},0) {$B'$};

\node[green!50!black,font=\small,below=8pt]
    at ({(-1.04275-1)/2},0) {$B'$};

\node[orange,font=\small,below=8pt]
    at ({(-3.98094-2.84275)/2},0) {$B$};

\node[blue,font=\small,below=8pt]
    at ({(1+1.04275)/2},0) {$A'$};

\node[blue,font=\small,below=8pt]
    at ({(6.58094+7.389056)/2},0) {$A'$};

\node[red,font=\small,below=8pt]
    at ({(2.84275+3.98094)/2},0) {$A$};

\foreach \x in {
    -7.389056,-6.58094,-3.98094,-2.84275,
    -1.04275,-1,
     1,1.04275,2.84275,3.98094,6.58094,7.389056
}{
    \fill (\x,0) circle (1.15pt);
}

\node[below=20pt,font=\scriptsize]
    at (-7.389056,0) {$-e^2$};

\node[below=20pt,font=\scriptsize]
    at (-6.58094,0) {$-6.581$};

\node[below=20pt,font=\scriptsize]
    at (-3.98094,0) {$-3.981$};

\node[below=20pt,font=\scriptsize]
    at (-2.84275,0) {$-2.843$};

\node[below=20pt,font=\scriptsize]
    at (-1.04275,0) {$-1.043$};

\node[below=20pt,font=\scriptsize]
    at (-1,0) {$-1$};

\node[below=20pt,font=\scriptsize]
    at (1,0) {$1$};

\node[below=20pt,font=\scriptsize]
    at (1.04275,0) {$1.043$};

\node[below=20pt,font=\scriptsize]
    at (2.84275,0) {$2.843$};

\node[below=20pt,font=\scriptsize]
    at (3.98094,0) {$3.981$};

\node[below=20pt,font=\scriptsize]
    at (6.58094,0) {$6.581$};

\node[below=20pt,font=\scriptsize]
    at (7.389056,0) {$e^2$};

\draw[black,line width=1.6pt]
    (2.535885,0.676898)
    arc[
        start angle=147.8256,
        end angle=50.2790,
        radius=1.2711745
    ];

\node[black,font=\small]
    at (3.56,1.43) {$H_A$};

\draw[black,line width=1.6pt]
    (6.419914,0.626687)
    arc[
        start angle=169.5508,
        end angle=142.3695,
        radius=3.4554105
    ];

\node[black,font=\small]
    at (6.73,2.05) {$H_{A'}$};

\draw[black,line width=1.6pt]
    (-4.424191,0.977743)
    arc[
        start angle=129.7210,
        end angle=32.1744,
        radius=1.2711745
    ];

\node[black,font=\small]
    at (-3.56,1.43) {$H_B$};

\draw[black,line width=1.6pt]
    (-7.081459,2.109761)
    arc[
        start angle=37.6305,
        end angle=10.4492,
        radius=3.4554105
    ];

\node[black,font=\small]
    at (-6.73,1.95) {$H_{B'}$};

\draw[black,line width=1.6pt]
    (1.154231,0.433864)
    arc[
        start angle=111.9092,
        end angle=142.3695,
        radius=0.4676390
    ];

\draw[black,line width=1.6pt]
    (-0.958371,0.285525)
    arc[
        start angle=37.6305,
        end angle=68.0908,
        radius=0.4676390
    ];

\draw[purple,dashed,line width=2.2pt]
    (4.32326,1.05345)
    arc[
        start angle=145.968,
        end angle=61.421,
        radius=1.88231
    ];

\draw[purple,dashed,line width=2.2pt]
    (-4.32326,1.05345)
    arc[
        start angle=34.032,
        end angle=118.579,
        radius=1.88231
    ];

\draw[purple,dashed,line width=2.2pt]
    (-3.16447,1.18982)
    arc[
        start angle=159.394,
        end angle=20.606,
        radius=3.38076
    ];

\draw[purple,dashed,line width=2.2pt]
    (1.161009,0.445961)
    arc[
        start angle=21.0125,
        end angle=158.9875,
        radius=1.243713
    ];

\draw[purple,dashed,line width=2.2pt]
    (6.422752,0.641906)
    arc[
        start angle=79.2942,
        end angle=80.4196,
        radius=0.653277
    ];

\draw[purple,dashed,line width=2.2pt]
    (-6.410121,0.644166)
    arc[
        start angle=99.5804,
        end angle=100.7058,
        radius=0.653277
    ];

\fill[purple] (4.32326,1.05345) circle (1.7pt);
\fill[purple] (6.78362,1.65296) circle (1.7pt);

\fill[purple] (-6.78362,1.65296) circle (1.7pt);
\fill[purple] (-4.32326,1.05345) circle (1.7pt);

\fill[purple] (-3.16447,1.18982) circle (1.7pt);
\fill[purple] (3.16447,1.18982) circle (1.7pt);

\fill[purple] (6.422752,0.641906) circle (1.7pt);
\fill[purple] (-6.422752,0.641906) circle (1.7pt);

\fill[purple]
    (1.161009,0.445961) circle (1.35pt);

\fill[purple]
    (-1.161009,0.445961) circle (1.35pt);

\fill[purple]
    (6.410121,0.644166) circle (1.35pt);

\fill[purple]
    (-6.410121,0.644166) circle (1.35pt);

\tikzset{
    htree/.style={
        htreecolor,
        double,
        double distance=1.2pt,
        line width=0.5pt
    }
}

\draw[htree]
    (5.02838461,1.67702324)
    arc[
        start angle=27.0084,
        end angle=91.2398,
        radius=3.692903
    ];

\draw[htree]
    (1.65832791,3.69203834)
    arc[
        start angle=151.2398,
        end angle=156.2227,
        radius=7.673437
    ];

\draw[htree]
    (1.65832791,3.69203834)
    arc[
        start angle=31.2398,
        end angle=75.5887,
        radius=7.118958
    ];

\draw[htree]
    (-0.3595375,0.9331306)
    arc[
        start angle=75.5887,
        end angle=79.6155,
        radius=0.963446
    ];

\draw[htree]
    (-3.57049588,6.46913501)
    arc[
        start angle=143.2320,
        end angle=170.1201,
        radius=10.807528
    ];

\draw[htree]
    (-0.48321407,0.87550221)
    arc[
        start angle=139.6155,
        end angle=143.2320,
        radius=1.462640
    ];

\draw[htree]
    (-0.4256560,0.9476652)
    arc[
        start angle=19.6155,
        end angle=23.7773,
        radius=2.822902
    ];

\fill[htreecolor]
    (1.65832791,3.69203834)
    circle (2.0pt);

\fill[htreecolor]
    (-0.4256560,0.9476652)
    circle (2.0pt);

\fill[htreecolor]
    (5.02838461,1.67702324)
    circle (1.5pt);

\fill[htreecolor]
    (1.36306564,3.09379996)
    circle (1.5pt);

\fill[htreecolor]
    (-5.56020653,1.85439189)
    circle (1.5pt);

\fill[htreecolor]
    (-0.5014435,1.1381455)
    circle (1.5pt);

\tikzset{
    htreeback/.style={
        backcolor,
        double,
        double distance=1.2pt,
        line width=0.5pt
    }
}

\draw[htreeback]
    (1.15705424,2.57602167)
    arc[
        start angle=40.3845,
        end angle=13.0546,
        radius=3.975870
    ];

\draw[htreeback]
    (5.19579751,1.29720887)
    arc[
        start angle=20.5651,
        end angle=27.0084,
        radius=3.692903
    ];

\draw[htreeback]
    (1.15705424,2.57602167)
    arc[
        start angle=160.3845,
        end angle=156.2227,
        radius=7.673437
    ];

\draw[htreeback]
    (1.15705424,2.57602167)
    arc[
        start angle=100.3845,
        end angle=148.7602,
        radius=2.618919
    ];

\draw[htreeback]
    (-0.61006463,1.35822566)
    arc[
        start angle=88.7602,
        end angle=140.5449,
        radius=1.358544
    ];

\draw[htreeback]
    (-5.64830270,1.24701433)
    arc[
        start angle=173.3742,
        end angle=170.1201,
        radius=10.807529
    ];

\draw[htreeback]
    (-0.61006463,1.35822566)
    arc[
        start angle=28.7602,
        end angle=23.7773,
        radius=2.822902
    ];

\fill[backcolor]
    (1.15705424,2.57602167)
    circle (2.0pt);

\fill[backcolor]
    (-0.61006463,1.35822566)
    circle (2.0pt);

\fill[backcolor]
    (2.00169478,0.89806765)
    circle (1.2pt);

\fill[backcolor]
    (5.19579751,1.29720887)
    circle (1.2pt);

\fill[backcolor]
    (-1.68842197,0.86331759)
    circle (1.2pt);

\fill[backcolor]
    (-5.64830270,1.24701433)
    circle (1.2pt);

\node[purple,font=\scriptsize]
    at (5.55,2.15)
    {$\mathrm{seam}(A,A')$};

\node[purple,font=\scriptsize]
    at (-5.55,2.15)
    {$\mathrm{seam}(B',B)$};

\node[purple,font=\scriptsize]
    at (0,3.65)
    {$\mathrm{seam}(B,A)$};

\node[purple,font=\scriptsize,xshift=10pt]
    at (0,1.47)
    {$\mathrm{seam}(A',B')$};

\node[
    brown!70!black,
    font=\small,
    fill=white,
    inner sep=2pt
]
    at (0,2.15)
    {back};

\node[
    orange!70!black,
    font=\small,
    fill=white,
    inner sep=2pt
]
    at (0,5.55)
    {front};

\node[
    htreecolor,
    above right,
    font=\scriptsize
]
    at (1.65832791,3.69203834)
    {$v_1$};

\node[
    htreecolor,
    below right,
       xshift=-11pt,
    yshift=-5pt,
    font=\scriptsize
]
    at (-0.4256560,0.9476652)
    {$v_2'=\gamma_1^{-1}(v_2)$};

\node[
    htreecolor,
    below,
    font=\scriptsize
]
    at (5.02838461,1.67702324)
    {$p_{AA'}$};

\node[
    htreecolor,
    left,
    font=\scriptsize
]
    at (1.36306564,3.09379996)
    {$p_{BA}$};

\node[
    htreecolor,
    below,
    font=\scriptsize
]
    at (-5.56020653,1.85439189)
    {$p_{B'B}$};

\node[
    htreecolor,
    below left,
    xshift=4pt,
    yshift=+6pt,
    font=\scriptsize
]
    at (-0.5014435,1.1381455)
    {$p_{A'B'}'$};

\node[
    backcolor,
    above left,
    font=\scriptsize
]
    at (1.15705424,2.57602167)
    {$v_3$};

\node[
    backcolor,
    above left,
    font=\scriptsize
]
    at (-0.61006463,1.35822566)
    {$v_4$};

\end{tikzpicture}

\caption{
Seam-anchored minimal H-trees at $\tau=0$, shown for the
\emph{$s$-channel} (Eq.~\eqref{eq:schannel}) on both front and back,
identically for $v_3,v_4$, at Parameter Choice~1 (Sec.~III):
the front-tree length is
$L^s_{\rm front}(0)=4.382590373\ldots$; the same qualitative structure
applies unchanged at Parameter Choice 2, where the corresponding
front-tree length is $L^s_{\rm front}(0)=5.233159625\ldots$ (Sec.~V),
although the dominant configuration there is instead the hybrid one (Table I). The red and blue double lines show the minimal $s$-channel networks on
the front and back halves, respectively.
The red network partitions the front half into the four regions
$A,A',B',B$.
The two trivalent vertices of the front network in the universal cover are
$v_1=(1.658328,3.692038)$ and
$v_2=(-3.145196,7.002351)$, where the three geodesic edges meet at
$120^\circ$.
Since $v_2$ lies outside the chosen Schottky fundamental region,
we display its image
$v_2'=\gamma_1^{-1}(v_2)=(-0.425656,0.947665)$;
similarly,
$p_{A'B'}$ is shown as
$p_{A'B'}'=\gamma_1^{-1}(p_{A'B'})$, and the geodesic
$\mathrm{seam}(A',B')$ itself is likewise shown via this same
$\gamma_1^{-1}$-translated lift.
Edges crossing the Schottky side identifications therefore appear
as disconnected pieces in the fundamental region, although on the
quotient each colored network forms a single five-edge H-tree.
As shown in Sec.~V, the competing $t$-channel network on the front
disk (not shown; it is simply the mirror image of the network
displayed here under $x\to-x$) is exactly degenerate with the network displayed
here, $L_{\rm front}^{t}(0)=L_{\rm front}^{s}(0)$; the seam anchors
$p_{AA'},p_{BA},p_{A'B'},p_{B'B}$ shown here are the values selected
by the joint front--back minimization of Eq.~\eqref{eq:Lglobal}, at
which the front and back legs meet each seam at a common point and
without a kink. (Figure prepared with the assistance of
Claude.)
}
\label{fig:htree-tau0}

\end{figure*}

\paragraph*{Seams and the front/back decomposition.}
It is useful to think of this construction in the spirit of cutting
a closed string into two open strings. The compact core of $\Sigma$ is a compact surface with boundary, and the four geodesic segments, the \emph{seams}
\begin{equation}
\begin{aligned}
\mathrm{seam}(A,A'),\quad
\mathrm{seam}(A',B'),\quad\\
\mathrm{seam}(B',B),\quad
\mathrm{seam}(B,A),
\end{aligned}
\end{equation}
cut it into two open pieces, the topological disks that we call
\emph{front} and \emph{back}. Each seam is the common perpendicular geodesic joining
the two horizons it separates, and its endpoints lie on those
horizons rather than on the asymptotic boundary. At $\tau=0$, front
and back are isometric, and each horizon of length $\ell$ is bisected
by its two adjacent seam feet into two arcs of length $\ell/2$ (see
Fig.~\ref{fig:htree-tau0}).

As derived in Appendix C, 
the hyperbolic distance
between two disjoint geodesics with ideal endpoints $0<r<s$ (after a
M\"obius transformation placing one of them on the imaginary axis)
is
\begin{equation}
d(\gamma_X,\gamma_Y)
=
\operatorname{arcosh}\!\left(\frac{r+s}{s-r}\right).
\label{eq:commonperp}
\end{equation}
Applying Eq.~\eqref{eq:commonperp} to the relevant pairs of horizon
lifts fixes the four seam lengths and their endpoints on the
fundamental domain.

\paragraph*{Topology of the front and back networks: two channels.}
Cutting each pair of pants along its three standard seams decomposes
it into two right-angled hexagons. Gluing the corresponding front
hexagons along their common half of the internal cuff $M$ produces
the front disk, whose boundary has cyclic order
\begin{equation}
\begin{aligned}
&A-\mathrm{seam}(A,A')-A'-\mathrm{seam}(A',B')-B'\\ & \qquad -\mathrm{seam}(B',B)-B-\mathrm{seam}(B,A) - A,
\end{aligned}
\end{equation}
and identically for the back disk. For this cyclic ordering, there are two non-crossing H-tree topologies, corresponding to the $s$- and $t$-channels of Ref.~\cite{Gadde:2023zzj} (their Fig.~17). We refer to these as the $s$-channel,
\begin{equation}
\begin{aligned}
v_1 &\sim\{\mathrm{seam}(A,A'),\mathrm{seam}(B,A)\}, \\
v_2 &\sim\{\mathrm{seam}(A',B'),\mathrm{seam}(B',B)\},
\label{eq:schannel}
\end{aligned}
\end{equation}
see Fig.~\ref{fig:htree-tau0},  and the $t$-channel,
\begin{equation}
\begin{aligned}
v_1 &\sim\{\mathrm{seam}(A,A'),\mathrm{seam}(A',B')\},\\
v_2 &\sim\{\mathrm{seam}(B,A),\mathrm{seam}(B',B)\},
\end{aligned}
\end{equation}
each joined by a single internal edge $v_1v_2$; the back disk admits
an identical pair of channels for its own trivalent vertices
$v_3,v_4$. Upon gluing the front and back disks, the corresponding
front and back legs are connected through their common seam anchors,
as described below. Explicitly, the front-network length in each channel is
\begin{equation}
\begin{aligned}
L_{\rm front}^{s} (v_1,v_2;\{p_i\}) &=d(v_1,p_{AA'})+d(v_1,p_{BA})+d(v_1,v_2)\\
&\quad+d(v_2,p_{A'B'})+d(v_2,p_{B'B}), \\
L_{\rm front}^{t} (v_1,v_2;\{p_i\})&=d(v_1,p_{AA'})+d(v_1,p_{A'B'})+d(v_1,v_2)\\
&\quad+d(v_2,p_{BA})+d(v_2,p_{B'B}),
\end{aligned}
\end{equation}
and $L_{\rm back}^{s},L_{\rm back}^{t}$ are defined identically for
$v_3,v_4$.

\paragraph*{From an open seam back to a closed global network.}
The seams are not physical boundaries of $\Sigma$: they are an
artificial cut, introduced solely to convert the closed-surface
Steiner problem on the compact core into two open, front/back
problems that are individually tractable as ordinary Steiner-tree
minimizations. Undoing this cut and gluing the two open pieces back
into the original closed surface requires the front and back
portions of the network to meet at the same seam point and to do so
smoothly. The seam anchors are shared variational variables of
the global problem: for each seam, the \emph{same} point $p_i$ must
be used by the front and back portions of the network. Extremizing
the total length with respect to $p_i$ imposes this
smooth-matching condition. As discussed below (``A direct-edge
shortcut''), when the corresponding direct geodesic crosses the
seam transversally in the appropriate homotopy class, the two
seam-mediated segments combine into a single smooth geodesic. Thus
the front and back networks cannot be minimized independently of
one another, even though the seam decomposition remains a
convenient way of parametrizing the global network. The H-tree
sector of the multi-way cut is described by the length
\begin{equation}
\begin{aligned}
&L_{\rm H}(\tau) \\
&=\min_{\substack{v_1,v_2,v_3,v_4\\ p_i\in\mathrm{seam}_i}}
\Big[L_{\rm front}(v_1,v_2;\{p_i\})+L_{\rm back}(v_3,v_4;\{p_i\})\Big]
\label{eq:Lglobal}
\end{aligned}
\end{equation}
minimized jointly over the four trivalent vertices, the four
\emph{shared} seam anchors $p_i$, and the discrete channel assignment
on each side \footnote{In contrast, an independent front/back minimization allows the front
and back portions to select different points on each seam and generically underestimates $L_{\rm H}$.}. $L_{\rm H}(\tau)$ is the resulting minimum within the H-tree sector.

Note that the true holographic
multi-entropy is obtained by minimizing over all admissible
configurations,
\begin{equation}
S^{(4)}(\tau)=\min_{\text{admissible configurations}} \frac{L(\tau)}{4G_N}.
\label{eq:S4global}
\end{equation}
Thus, for the H-tree constructed above to determine $S^{(4)}$, it must also win the global competition against other admissible configurations. The competing sectors analyzed below are the H-tree, branchless, and
hybrid configurations,
\begin{equation}
L_{\rm H}(\tau),\qquad L_{\rm branchless},\qquad L_{\rm hybrid},
\end{equation}
the latter two being exactly $\tau$-independent, as shown below. The
full $S^{(4)}$ is obtained by comparing these in Sec.~V.

\paragraph*{A direct-edge shortcut.}
Because the horizon associated with each seam endpoint is identified
with its Schottky-paired image under $\gamma_2$, $\gamma_1$, or
$\gamma_2'(\tau)$ (Sec.~II), extremizing Eq.~\eqref{eq:Lglobal} over
each shared anchor $p_i$, for fixed $v_1,v_2,v_3,v_4$, is equivalent
to minimizing $d(v,p_i)+d(p_i,v')$ over $p_i$ constrained to the
seam, where $v$ and $v'$ are the
corresponding front and back trivalent vertices, with $v'$ mapped by
the appropriate Schottky side-pairing transformation when necessary. In
general this constrained minimum only satisfies
\begin{equation}
d(v,p_i)+d(p_i,v')\geq d(v,v'), 
\end{equation}
with equality iff the geodesic segment joining $v$ and $v'$ directly
crosses the seam in the correct homotopy class. We call such a crossing \emph{transversal} when the connecting geodesic crosses, rather than merely touches, the seam.

At $\tau=0$, we find numerically that this crossing is orthogonal
(the crossing angle equals $\pi/2$ to the precision of our
optimization). The exact value of the angle is not needed below:
what matters is only that the crossing is transversal, namely that
the connecting geodesic crosses the seam rather than becoming
tangent to it. Since the minimizing configuration varies smoothly
with $\tau$ near $\tau=0$, transversality persists for sufficiently
small $|\tau|$; we further verify numerically that it persists
throughout the range of $\tau$ studied below, so that
Eq.~\eqref{eq:Lglobal} may be evaluated, to within our numerical
precision, using the direct edges
\begin{equation}
\begin{aligned}
&\, d_{AA'}(v,v')\equiv d\big(v,\gamma_2^{-1}v'\big), \,\,\,\,
d_{BA}(v,v')\equiv d(v,v'), \\
&\, \hspace{-1mm}d_{A'B'}(v,v')\equiv d\big(v,\gamma_1v'\big), \quad
d_{B'B}(v,v')\equiv d\big(v,\gamma_2'(\tau)^{-1}v'\big),
\end{aligned}
\label{eq:folds}
\end{equation}
in place of each pair of seam-mediated segments. We stress that Eq.~\eqref{eq:folds} is used only as a
computationally efficient shortcut for evaluating $L_{\rm H}$,
valid precisely because transversality holds; it is $L_{\rm H}$ of Eq.~\eqref{eq:Lglobal} that
is being minimized, and the shortcut Eq.~\eqref{eq:folds} would
simply fail to reproduce it wherever transversality broke down.

If front and back are assigned the \emph{same} channel (both $s$ or
both $t$), $v_1$ and $v_3$ are then joined by two edges---via
$\mathrm{seam}(A,A')$ and $\mathrm{seam}(B,A)$---and $v_2,v_4$
likewise by two edges; if front and back are assigned
\emph{different} channels, the four edges instead connect
$v_1v_3,v_1v_4,v_2v_3,v_2v_4$ once each, so that together with
$v_1v_2,v_3v_4$ the graph is the complete graph $K_4$. Both graphs are trivalent interface graphs compatible with a
four-region partition of the compact core, so there is no
topological obstruction to either. Which is shorter is a dynamical question, settled in Sec.~V in favor of the same-channel graph.

\section{V. $\tau=0$ and Its Neighborhood: The Global Minimum}

We now show the results for Parameter Choice 1 and 2.

\subsection*{Parameter Choice~1}

\paragraph*{Trivalent H-tree Steiner configurations:}

At $\tau=0$, we first verify a single-sided fact: minimizing the
front network alone in either channel gives an exact degeneracy. 
We find at Parameter Choice~1: \begin{equation}
L_{\rm front}^{s}(0)=L_{\rm
front}^{t}(0)=4.382590373\ldots ,
\end{equation}
to within numerical precision, both trivalent vertices satisfying
the $120^\circ$ Steiner condition and each leg meeting its seam
orthogonally. See Fig.~\ref{fig:htree-tau0}.

This is not a coincidence: it follows from the
$\mathbb{Z}_2$ symmetry of Sec.~II, $c_1'=-c_1,c_2'=-c_2$, under
which $x\to-x$ together with $A\leftrightarrow B$,
$A'\leftrightarrow B'$ maps the minimizing $s$-channel configuration
onto the minimizing $t$-channel one. This reflection descends to an
isometry of the quotient geometry only at $\tau=0$, and this is common to Parameter Choice 1 and 2. For
$\tau\neq0$, it conjugates $\gamma_2'(\tau)$ to
$\eta(\tau)\gamma_2\eta(-\tau)$ rather than to $\gamma_2$ itself, and thus this isometry is broken at $\tau \neq 0$.

Assembling the front and back networks for the same-channel
(e.g.\ front $s$, back $s$), we find 
\begin{equation}
L_{\rm H}^{\rm same}(0) =8.765180746\ldots .
\end{equation}
For mixed-channel (e.g.\ front $s$, back
$t$; $K_4$) graphs, we find 
\begin{equation}
L_{\rm H}^{\rm mixed}(0)=9.050969924\ldots.
\end{equation}
Thus, for trivalent H-tree Steiner configurations, the same-channel dominates.

\paragraph*{Response of the H-tree to the twist:}
Although, as we show below, this H-tree sector is not the global
minimum at either parameter choice studied in this paper, its own
response to the twist is a well-defined property worth recording \footnote{The value $L_{\rm H}^{\rm same}(0)$ coincides with that
obtained by minimizing the front and back seam anchors
{independently} of one another, omitting the matching
condition that a shared, kink-free anchor must satisfy. This is a
consequence of the $\tau=0$ symmetry: the independently-optimized
front and back anchors on each seam happen to coincide at this
exactly symmetric point (verified numerically to seven significant
figures), so the matching condition is automatically satisfied
without being imposed. We find numerically that this is special to
$\tau=0$: for $\tau\neq0$ the independently-optimized anchors no
longer coincide, and enforcing the matching condition is essential
to obtain the correct $L_{\rm H}(\tau)$.}. 
For $\tau\neq0$, the twist is implemented on the second pair of
pants only, via $\gamma_2'(\tau)=\eta(\tau)\gamma_2'\eta(-\tau)$
(Sec.~II); consequently the horizon $H_A,H_{A'}$ and the seam
$\mathrm{seam}(A,A')$ are exactly $\tau$-independent, while
$H_B,H_{B'}$ and the remaining three seams move with $\tau$.

Writing $L_{\rm H}(\tau)=\min_x F(x;\tau)$ for the joint
optimization variables $x=(v_1,v_2,v_3,v_4, \{p_i\})$, the envelope theorem
gives $L'_{\rm H}(0)=\partial_\tau F(x^*;0)|_{x=x^*(0)}$, requiring
no explicit re-optimization of $x$. We find $L'_{\rm H}(0)=0$
exactly, not as an approximate cancellation between two
independently-computed slopes but as a direct consequence of an
exact symmetry: the front--back exchange combined with
$A\leftrightarrow B$ acts on the joint network as $\tau\to-\tau$, so
that $L_{\rm H}(\tau)=L_{\rm H}(-\tau)$ identically, and in
particular $L_{\rm H}$ is an even function of $\tau$ with no linear
term. We have confirmed this evenness numerically to ten significant
figures over the range $|\tau|\le0.2$ \footnote{Strictly speaking,
evenness alone does not exclude a nonanalytic cusp proportional to
$|\tau|$ at $\tau=0$. In the present case, however, the numerically
minimized branch continues smoothly through $\tau=0$: $[L_{\rm
H}(\tau)-L_{\rm H}(0)]/\tau^2$ converges to a finite constant as
$\tau\to0$, while $[L_{\rm H}(\tau)-L_{\rm H}(0)]/|\tau|\to0$. Hence
no cusp is observed and $L'_{\rm H}(0)=0$.}.

At second order, writing the stationarity condition
$\nabla_xF(x^*(\tau),\tau)=0$ and differentiating in $\tau$ gives
$x^{*\prime}(0)=-H^{-1}g$, with $H=\nabla_x^2F(x^*,0)$ and
$g=\nabla_x\partial_\tau F(x^*,0)$, so that $L''_{\rm
H}(0)=F_{\tau\tau}(x^*,0)-g^TH^{-1}g$. We evaluate this expression
numerically using the explicit shared-anchor parametrization, and
independently confirm it using the optimized folded formulation of
Eq.~\eqref{eq:folds} with central differences and Richardson
extrapolation; the two methods, sharing no intermediate steps,
agree to five significant figures, giving
\begin{equation}
L''_{\rm H}(0)\approx0.5956 \quad\text{at Parameter Choice~1.}
\end{equation}
Combining the two orders, $L_{\rm H}(\tau)-L_{\rm
H}(0)\approx(0.5956/2)\,\tau^2+O(\tau^4)$ as $\tau\to0$: the H-tree
sector itself is twist-sensitive.

\paragraph*{Branchless configurations:}

In addition to these trivalent H-tree configurations, the compact core
also admits \emph{branchless} configurations, {\it i.e.,} unions of mutually
disjoint simple closed geodesics, with no trivalent junction, that separate the compact core into four
regions, one containing each of $A,A',B,B'$. Any branchless
configuration containing a closed geodesic that crosses $M$ is
excluded by the bound $L\geq2z$ of Sec.~III: completing such a curve
to a four-region partition requires two additional horizon components,
giving a total length of at least $2z+2\ell$, which is
$11.224452\ldots$ at Parameter Choice~1, far above every candidate
considered below. On the other hand, any essential simple closed
geodesic disjoint from $M$ is boundary-parallel within one of the
two pairs of pants. Thus the only remaining building blocks are the
four horizons and the internal cuff $M$. An exhaustive check of the
$\binom{5}{3}=10$ ways of selecting three of these five curves shows
that the valid four-region partitions have only two types: three
horizons, of length $3\ell$, or $M$ together with one horizon from
each pair of pants, of length $2\ell+m$. Both configurations satisfy
the required homology and domain-partition conditions for $S^{(4)}$
and must therefore be included as admissible competitors alongside
the H-tree networks. On our symmetric parameter choice $\ell=m$ studied here,
these two branchless lengths coincide, and we have
\begin{equation}
L_{\rm branchless}=3\ell=2\ell+m = 6,
\end{equation}
at Parameter Choice~1.

\paragraph*{Hybrid configurations:}
Cutting the compact core along $M$ separates it into its two
constituent pairs of pants, $\{A,A',M\}$ and $\{M, B,B'\}$. A further
admissible configuration treats these two pieces asymmetrically:
the pair of pants $\{A,A',M\}$ is equipped with its own front/back
Steiner tripod, while the pair of pants $\{M, B,B'\}$ is simply
cut along horizon $B$ (or, equivalently since $L_B=L_{B'}$,
horizon $B'$), with no trivalent vertex on that side. 
The remaining piece of $\{M,B,B'\}$, bounded by $M$ and horizon $B$,
still reaches the $B'$ funnel; upon gluing along $M$, it joins
smoothly with the $M$-facing wedge of the $\{A,A',M\}$ tripod to
form a single connected region containing only the $B'$ boundary. The result is a
valid four-region partition, distinct from both the H-tree and the
branchless configurations above.

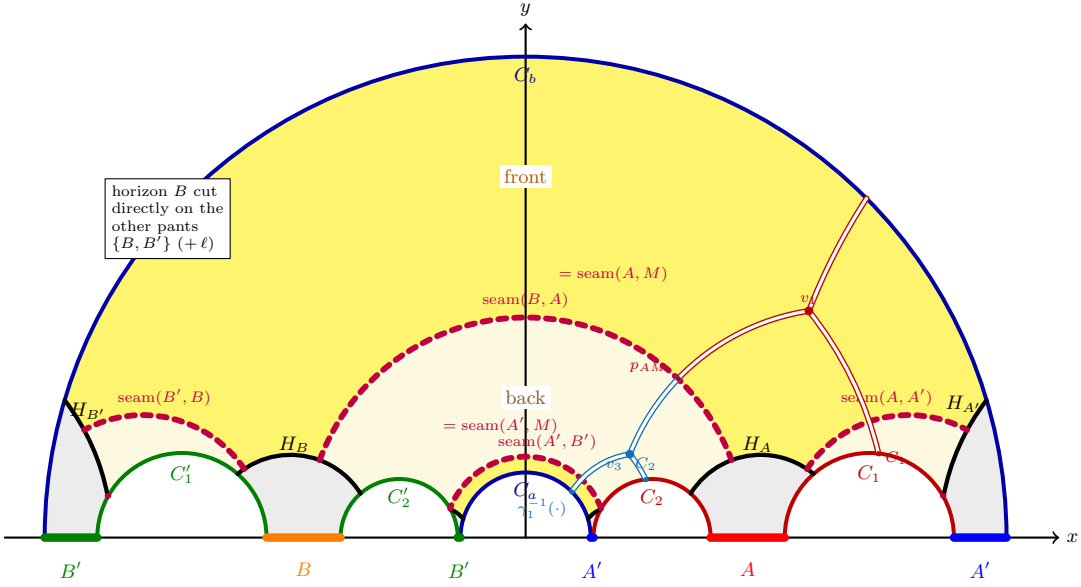
\begin{figure*}[t]
\centering

\begin{tikzpicture}[
    x=1.05cm,
    y=1.05cm,
    scale=0.82,
    transform shape,
    line cap=round,
    line join=round
]

\def\muR{7.389056}
\def\cOne{5.28094}
\def\rOne{1.3}
\def\cTwo{1.94275}
\def\rTwo{0.9}

\definecolor{htreecolor}{RGB}{200,20,40}
\definecolor{backcolor}{RGB}{20,110,190}

\fill[yellow!70]
    (-\muR,0)
    arc[start angle=180,end angle=0,radius=\muR]
    -- cycle;

\fill[white]
    (-1,0)
    arc[start angle=180,end angle=0,radius=1]
    -- cycle;

\fill[white]
    ({\cOne-\rOne},0)
    arc[start angle=180,end angle=0,radius=\rOne]
    -- cycle;

\fill[white]
    ({\cTwo-\rTwo},0)
    arc[start angle=180,end angle=0,radius=\rTwo]
    -- cycle;

\fill[white]
    ({-\cTwo-\rTwo},0)
    arc[start angle=180,end angle=0,radius=\rTwo]
    -- cycle;

\fill[white]
    ({-\cOne-\rOne},0)
    arc[start angle=180,end angle=0,radius=\rOne]
    -- cycle;

\fill[gray!15]
    (2.84275,0) -- (3.98094,0)
    arc[start angle=180,end angle=131.2265,radius=1.3]
    arc[start angle=50.2790,end angle=147.8256,radius=1.2711745]
    arc[start angle=48.7734,end angle=0,radius=0.9]
    -- cycle;

\fill[gray!15]
    (-3.98094,0) -- (-2.84275,0)
    arc[start angle=180,end angle=131.2266,radius=0.9]
    arc[start angle=32.1744,end angle=129.7210,radius=1.2711745]
    arc[start angle=48.7735,end angle=0,radius=1.3]
    -- cycle;

\fill[gray!15]
    (6.58094,0) -- (7.389056,0)
    arc[start angle=0,end angle=16.5902,radius=7.389056]
    arc[start angle=142.3695,end angle=169.5508,radius=3.4554105]
    arc[start angle=28.8205,end angle=0,radius=1.3]
    -- cycle;

\fill[gray!15]
    (1,0) -- (1.04275,0)
    arc[start angle=180,end angle=151.1792,radius=0.9]
    arc[start angle=111.9092,end angle=142.3695,radius=0.4676390]
    arc[start angle=16.5902,end angle=0,radius=1]
    -- cycle;

\fill[gray!15]
    (-7.389056,0) -- (-6.58094,0)
    arc[start angle=180,end angle=151.1795,radius=1.3]
    arc[start angle=10.4492,end angle=37.6305,radius=3.4554105]
    arc[start angle=163.4098,end angle=180,radius=7.389056]
    -- cycle;

\fill[gray!15]
    (-1.04275,0) -- (-1,0)
    arc[start angle=180,end angle=163.4098,radius=1]
    arc[start angle=37.6305,end angle=68.0908,radius=0.4676390]
    arc[start angle=28.8208,end angle=0,radius=0.9]
    -- cycle;

\fill[yellow!75!brown!15]
    (2.535885,0.676898)
    arc[start angle=147.8256,end angle=110.6065,radius=1.2711745]
    arc[start angle=20.606,end angle=159.394,radius=3.38076]
    arc[start angle=69.3935,end angle=32.1744,radius=1.2711745]
    arc[start angle=131.2266,end angle=29.7035,radius=0.9]
    arc[start angle=158.9875,end angle=21.0125,radius=1.243713]
    arc[start angle=150.2965,end angle=48.7734,radius=0.9]
    -- cycle;

\fill[yellow!75!brown!15]
    (4.424191,0.977743)
    arc[start angle=50.2790,end angle=55.9682,radius=1.2711745]
    arc[start angle=145.968,end angle=61.421,radius=1.88231]
    arc[start angle=151.4211,end angle=169.2941,radius=3.4554105]
    arc[start angle=79.2942,end angle=80.4196,radius=0.653277]
    arc[start angle=29.7035,end angle=131.2265,radius=1.3]
    -- cycle;

\fill[yellow!75!brown!15]
    (-4.424191,0.977743)
    arc[start angle=129.7210,end angle=124.0318,radius=1.2711745]
    arc[start angle=34.032,end angle=118.579,radius=1.88231]
    arc[start angle=28.5789,end angle=10.7059,radius=3.4554105]
    arc[start angle=100.7058,end angle=99.5804,radius=0.653277]
    arc[start angle=150.2965,end angle=48.7735,radius=1.3]
    -- cycle;

\draw[->,line width=0.8pt] (-8.0,0) -- (8.2,0) node[right] {$x$};
\draw[->,line width=0.8pt] (0,0) -- (0,7.9) node[above] {$y$};

\draw[blue!65!black,line width=1.5pt] (-\muR,0) arc[start angle=180,end angle=0,radius=\muR];
\draw[blue!65!black,line width=1.5pt] (-1,0) arc[start angle=180,end angle=0,radius=1];
\draw[red!75!black,line width=1.5pt] ({\cOne-\rOne},0) arc[start angle=180,end angle=0,radius=\rOne];
\draw[red!75!black,line width=1.5pt] ({\cTwo+\rTwo},0) arc[start angle=0,end angle=180,radius=\rTwo];
\draw[green!50!black,line width=1.5pt] ({-\cOne+\rOne},0) arc[start angle=0,end angle=180,radius=\rOne];
\draw[green!50!black,line width=1.5pt] ({-\cTwo-\rTwo},0) arc[start angle=180,end angle=0,radius=\rTwo];

\node[blue!65!black,font=\small] at (0,7.10) {$C_b$};
\node[blue!65!black,font=\small] at (0,0.72) {$C_a$};
\node[red!75!black,font=\small] at (\cOne,0.98) {$C_1$};
\node[red!75!black,font=\small] at (\cTwo,0.62) {$C_2$};
\node[green!50!black,font=\small] at (-\cTwo,0.62) {$C_2'$};
\node[green!50!black,font=\small] at (-\cOne,0.98) {$C_1'$};

\draw[green!50!black,line width=3.2pt] (-7.389056,0) -- (-6.58094,0);
\draw[green!50!black,line width=3.2pt] (-1.04275,0) -- (-1,0);
\draw[orange,line width=3.2pt] (-3.98094,0) -- (-2.84275,0);
\draw[blue,line width=3.2pt] (1,0) -- (1.04275,0);
\draw[blue,line width=3.2pt] (6.58094,0) -- (7.389056,0);
\draw[red,line width=3.2pt] (2.84275,0) -- (3.98094,0);

\node[green!50!black,font=\small,below=8pt] at ({(-7.389056-6.58094)/2},0) {$B'$};
\node[green!50!black,font=\small,below=8pt] at ({(-1.04275-1)/2},0) {$B'$};
\node[orange,font=\small,below=8pt] at ({(-3.98094-2.84275)/2},0) {$B$};
\node[blue,font=\small,below=8pt] at ({(1+1.04275)/2},0) {$A'$};
\node[blue,font=\small,below=8pt] at ({(6.58094+7.389056)/2},0) {$A'$};
\node[red,font=\small,below=8pt] at ({(2.84275+3.98094)/2},0) {$A$};

\draw[black,line width=1.6pt] (2.535885,0.676898) arc[start angle=147.8256,end angle=50.2790,radius=1.2711745];
\node[black,font=\small] at (3.56,1.43) {$H_A$};
\draw[black,line width=1.6pt] (6.419914,0.626687) arc[start angle=169.5508,end angle=142.3695,radius=3.4554105];
\node[black,font=\small] at (6.73,2.05) {$H_{A'}$};
\draw[black,line width=1.6pt] (-4.424191,0.977743) arc[start angle=129.7210,end angle=32.1744,radius=1.2711745];
\node[black,font=\small] at (-3.56,1.43) {$H_B$};
\draw[black,line width=1.6pt] (-7.081459,2.109761) arc[start angle=37.6305,end angle=10.4492,radius=3.4554105];
\node[black,font=\small] at (-6.73,1.95) {$H_{B'}$};
\draw[black,line width=1.6pt] (1.154231,0.433864) arc[start angle=111.9092,end angle=142.3695,radius=0.4676390];
\draw[black,line width=1.6pt] (-0.958371,0.285525) arc[start angle=37.6305,end angle=68.0908,radius=0.4676390];

\draw[purple,dashed,line width=2.2pt] (4.32326,1.05345) arc[start angle=145.968,end angle=61.421,radius=1.88231];
\draw[purple,dashed,line width=2.2pt] (-4.32326,1.05345) arc[start angle=34.032,end angle=118.579,radius=1.88231];
\draw[purple,dashed,line width=2.2pt] (-3.16447,1.18982) arc[start angle=159.394,end angle=20.606,radius=3.38076];
\draw[purple,dashed,line width=2.2pt] (1.161009,0.445961) arc[start angle=21.0125,end angle=158.9875,radius=1.243713];
\draw[purple,dashed,line width=2.2pt] (6.422752,0.641906) arc[start angle=79.2942,end angle=80.4196,radius=0.653277];
\draw[purple,dashed,line width=2.2pt] (-6.410121,0.644166) arc[start angle=99.5804,end angle=100.7058,radius=0.653277];

\node[purple,font=\scriptsize] at (5.55,2.15) {$\mathrm{seam}(A,A')$};
\node[purple,font=\scriptsize] at (-5.55,2.15) {$\mathrm{seam}(B',B)$};
\node[purple,font=\scriptsize] at (0,3.65) {$\mathrm{seam}(B,A)$};
\node[purple,font=\scriptsize,xshift=10pt] at (0,1.47) {$\mathrm{seam}(A',B')$};

\node[brown!70!black,font=\small,fill=white,inner sep=2pt] at (0,2.15) {back};
\node[orange!70!black,font=\small,fill=white,inner sep=2pt] at (0,5.55) {front};

\node[purple,font=\scriptsize] at (1.35,4.05) {$=\mathrm{seam}(A,M)$};
\node[purple,font=\scriptsize,xshift=-4pt] at (-0.25,1.72) {$=\mathrm{seam}(A',M)$};

\tikzset{fronttripod/.style={red!75!black,double,double distance=1.2pt,line width=0.5pt}}
\tikzset{backtripod/.style={backcolor,double,double distance=1.2pt,line width=0.5pt}}

\draw[fronttripod]
   ({4.883022+3.5234*cos(98.674)},{3.5234*sin(98.674)})
   arc[start angle=98.674,end angle=136.184,radius=3.5234];
\draw[backtripod]
   ({4.883022+3.5234*cos(136.184)},{3.5234*sin(136.184)})
   arc[start angle=136.184,end angle=158.674,radius=3.5234];

\draw[fronttripod]
   ({5.573936*cos(38.674)},{5.573936*sin(38.674)})
   arc[start angle=38.674,end angle=13.406,radius=5.573936];
\draw[backtripod]
   ({2.050536*cos(25.869)},{2.050536*sin(25.869)})
   arc[start angle=25.869,end angle=38.674,radius=2.050536];

\draw[fronttripod]
   ({13.273431+9.577594*cos(158.674)},{9.577594*sin(158.674)})
   arc[start angle=158.674,end angle=147.033,radius=9.577594];
\draw[backtripod]
   ({1.796364+1.296186*cos(147.033)},{1.296186*sin(147.033)})
   arc[start angle=147.033,end angle=98.674,radius=1.296186];

\fill[red!75!black] (4.351642,3.483099) circle (2.0pt);
\node[red!75!black,font=\scriptsize,above] at (4.351642,3.483099) {$v_1$};

\fill[backcolor] (1.600880,1.281361) circle (2.0pt);
\node[backcolor,font=\scriptsize,below left] at (1.600880,1.281361) {$v_3$};

\fill[red!75!black] (5.422055,1.292319) circle (1.1pt);
\node[red!75!black,font=\scriptsize,right,yshift=-2pt] at (5.422055,1.292319) {$C_1$};

\fill[backcolor] (1.845059,0.894682) circle (1.1pt);
\node[backcolor,font=\scriptsize,above,yshift=1pt] at (1.845059,0.894682) {$C_2$};

\fill[red!75!black] (2.340673,2.439428) circle (1.1pt);
\node[red!75!black,font=\scriptsize,above left,xshift=-2pt] at (2.340673,2.439428) {$p_{AM}$};

\fill[backcolor] (0.708883,0.705326) circle (1.1pt);
\node[backcolor,font=\scriptsize,below left,xshift=1pt] at (0.708883,0.705326) {$\gamma_1^{-1}(\cdot)$};

\node[font=\scriptsize,align=left,draw,fill=white,inner sep=3pt] at (-5.5,4.9)
   {horizon $B$ cut\\ directly on the\\ other pants\\ $\{B,B'\}$ ($+\,\ell$)};

\end{tikzpicture}

\caption{
The hybrid configuration at Parameter Choice~1 ($\ell=m=2$), shown
on the same Schottky fundamental region as
Fig.~\ref{fig:htree-tau0}. The pair of pants $\{A,A',M\}$ contains
a Steiner tripod with distinct front and back trivalent vertices
$v_1$ (red) and $v_3$ (blue), connected across the three seams as
in Sec.~IV, while the other pair of pants is cut along horizon $B$.
The tripod has total length $4.553522553\ldots$; including horizon
$B$ gives
$L_{\rm hybrid}=6.553522553\ldots$.
(Figure prepared with the assistance of Claude.)
}
\label{fig:hybrid}
\end{figure*}

Explicitly, the front and back tripods of the $\{A,A',M\}$ pair of pants are each a Steiner tree reaching the three shared anchors on
$\mathrm{seam}(A,A')$, $\mathrm{seam}(A,M)$ (= half of $\mathrm{seam}(B, A)$), and $\mathrm{seam}(A',M)$ (= half of $\mathrm{seam}(A',B')$),
the latter two being common perpendiculars between the corresponding
horizon and the $M$-lift $\widetilde M$, computable directly from
Eq.~\eqref{eq:commonperp}. At $\tau=0$, front and back are isometric, as in Sec.~IV, so the two tripods
have equal length, and
\begin{equation}
\begin{aligned}
L_{\rm hybrid}&=
2\min_{v,\,p_{AA'},p_{AM},p_{A'M}}
\Big[d(v,p_{AA'})  \\
&   \qquad \qquad +d(v,p_{AM}) +d(v,p_{A'M})\Big]  +\ell,
\label{eq:Lhybrid}
\end{aligned}
\end{equation}
where the final $+\ell$ is the length of horizon $B$ (or $B'$) itself. 
See Fig.~\ref{fig:hybrid}. Since
$\mathrm{seam}(A,A')$, $\mathrm{seam}(A,M)$, and $\mathrm{seam}(A',M)$
are all built from $\gamma_2$, $\gamma_1$, and the fixed $M$-lift
alone, none of which depend on $\tau$, and since the length of
horizon $B$ is exactly $\tau$-independent by trace invariance
(Sec.~II) regardless of its position, $L_{\rm hybrid}$ is exactly
$\tau$-independent for \emph{all} $\tau$, not merely at $\tau=0$.

At Parameter Choice~1, Eq.~\eqref{eq:Lhybrid} gives
\begin{equation}
L_{\rm hybrid}
=2\times2.276761276\ldots \, +\, \ell
=6.553522553\ldots.
\end{equation}

\paragraph*{Dominant configuration:}
Collecting all these results, at Parameter Choice~1, we have
\begin{equation}
L_{\rm branchless}<L_{\rm hybrid}<L_{\rm H}^{\rm same}(0)<L_{\rm H}^{\rm mixed}(0),
\end{equation}
so, within the class of admissible configurations identified and
analyzed here (the same-channel and mixed-channel H-trees, the
branchless configurations, and the hybrid configuration, together
with the curves crossing $M$ excluded above for exceeding the bound $2z$), we find the
branchless configuration to be the global minimum.   Being built from horizon and cuff geodesics alone, it is exactly $\tau$-independent, and thus, $S^{(4)}$ does not detect the twist at Parameter Choice~1.

This is the central lesson of Parameter Choice~1: the existence of a
connected H-tree does not by itself imply that $S^{(4)}$ is
determined by that configuration, since $S^{(4)}$ is given by a
global minimization (Sec.~IV) that may instead select a
$\tau$-independent branchless competitor. This raises the question
of whether a parameter choice exists at which the connected H-tree
instead wins the global competition, while the bipartite RT data
remain exactly $\tau$-independent as required by Sec.~III.

\subsection*{Parameter Choice~2}

\paragraph*{Trivalent H-tree Steiner configurations:}

Similar to Parameter Choice 1, 
by the same $\mathbb{Z}_2$ symmetry,
\begin{equation}
L_{\rm front}^{s}(0)=L_{\rm
front}^{t}(0)=5.233159625\ldots 
\end{equation}
Thus, we find 
\begin{equation}
L_{\rm H}^{\rm same}(0)=10.466319251\ldots,
\end{equation}
Similarly for mixed-channel, we find
\begin{equation}
L_{\rm H}^{\rm mixed}(0)=10.602886951\ldots.
\end{equation}
Thus, for trivalent H-tree Steiner configurations, the same-channel
dominates.

\paragraph*{Response of the H-tree to the twist:}
By the same method as at Parameter Choice~1, we find
\begin{equation}
L'_{\rm H}(0)=0,\qquad L''_{\rm H}(0)\approx0.4614,
\end{equation}
the two independent methods again agreeing to five significant
figures, so that $L_{\rm H}(\tau)-L_{\rm H}(0)\approx
0.2307\,\tau^2+O(\tau^4)$ as $\tau\to0$.

\paragraph*{Branchless configurations:}
The branchless competitor, evaluated by the same argument as above
(with the exclusion bound for curves crossing $M$ now
$2z+2\ell=14.384848920\ldots$ at this point), gives 
\begin{equation}
L_{\rm
branchless}=3\ell=2\ell+m=10.65
\end{equation}
at Parameter Choice~2.

\paragraph*{Hybrid configurations:}
For the
hybrid configuration, the identical construction at Parameter Choice~2 gives
\begin{equation}
L_{\rm hybrid}
=2\times3.235297970\ldots+\ell
=10.020595940\ldots.
\end{equation}

\paragraph*{Dominant configuration:}
Collecting these,
\begin{equation}
L_{\rm hybrid}<L_{\rm H}^{\rm same}(0)<L_{\rm H}^{\rm mixed}(0)<L_{\rm branchless},
\end{equation}
so within the same class of competitors, the hybrid configuration is the global minimum at Parameter Choice~2. Being exactly
$\tau$-independent, 
$S^{(4)}$ does not detect the twist at
Parameter Choice~2 either.

\subsection*{Summary for small $\tau$}
Collecting these results (Table~\ref{tab:choices}):
\begin{table}[h]
\centering
\begin{tabular}{lcc}
\hline\hline
 & Choice 1 ($\ell=m=2$) & Choice 2 ($\ell=m=3.55$) \\
\hline
$L_{\rm H}^{\rm same}$        & $8.765180746\ldots$  & $10.466319251\ldots$ \\
$L_{\rm H}^{\rm mixed}$       & $9.050969924\ldots$  & $10.602886951\ldots$ \\
$L_{\rm branchless}$  & $6$                  & $10.65$              \\
$L_{\rm hybrid}$      & $6.553522553\ldots$  & $10.020595940\ldots$ \\
\hline
dominant              & branchless           & hybrid               \\
\hline\hline
\end{tabular}
\caption{Comparison of the competing configurations considered here, at the two parameter choices, at $\tau=0$, among which the tabulated minimum is taken. }
\label{tab:choices}
\end{table}

Among these four competitors, $L_{\rm branchless}$ and $L_{\rm
hybrid}$ are exactly $\tau$-independent, while the H-trees are
generically $\tau$-dependent connected networks; we have explicitly
computed the twist response only for the same-channel graph. Since
the values in Table~\ref{tab:choices} show finite gaps, continuity
ensures the comparison persists for sufficiently small $|\tau|$.

\section{VI. Varying $\ell=m$: Which Configuration Wins?}

Table~\ref{tab:choices} compares the competing configurations at two
representative points. To understand how this competition unfolds
more broadly, we now fix $\ell=m$ throughout \footnote{\label{fn:closedform}Once $\ell,m$ are fixed, the pair of pants $\{A,A',M\}$ is
rigid, so $(c_1,c_2,R_1,R_2)$ merely coordinatize this unique shape;
different valid choices are related by conjugation and give
identical isometry-invariant quantities ($z$, $L_{\rm H}$, $L_{\rm
hybrid}$, etc.), which we have verified numerically to twelve
significant figures. We are therefore free to choose $(R_1,R_2)$ at
each $\ell$ however is convenient, provided the six Schottky
circles remain disjoint; holding $(R_1,R_2)$ fixed at a single
value throughout, for instance, does not remain valid over the full
range. Using the exact closed-form solution of the trace equations,
$c_1=\mu D/(\mu-1)$, $c_2=D/(\mu-1)$, with
$D\equiv2\sqrt{R_1R_2}\cosh(\ell/2)$ (valid on the branch
$c_1-c_2=D$, $c_1/\mu-c_2\mu=-D$, reducing to the values of Sec.~III
at $\ell=2$), together with the explicit gauge $(R_1,R_2)=
(1.3+\frac{\ell-2}{1.55}\times4.55,\ 0.9-\frac{\ell-2}{1.55}\times0.70)$
for $2\le\ell\le3.55$ and $(5.85,0.20)$ for $\ell\ge3.55$, we have
verified disjointness, Eq.~\eqref{parameterconsistency}, directly
and exactly for $\ell=m$ ranging from $2$ to $4$, with the tightest
margin, $c_2-R_2-1\approx0.043$, occurring at $\ell=2$ itself.}, along the
symmetric line of Sec.~III, and track how $L_{\rm H}^{\rm same}$,
$L_{\rm branchless}=3\ell$, and $L_{\rm hybrid}$ move against one
another as $\ell$ increases from Parameter Choice~1.

\paragraph*{$\ell\lesssim2.66$.}
At $\ell=2$ (Parameter Choice~1),
\begin{equation}
L_{\rm branchless}<L_{\rm hybrid}<L_{\rm H}^{\rm same},
\end{equation}
and this ordering persists as $\ell$ increases, up to $\ell\approx2.66$.

\paragraph*{$2.66\lesssim\ell\lesssim3.43$.}
At $\ell\approx2.66$, $L_{\rm hybrid}$ and $L_{\rm branchless}$
cross, and the ordering becomes
\begin{equation}
L_{\rm hybrid}<L_{\rm branchless}<L_{\rm H}^{\rm same}.
\end{equation}

\paragraph*{$3.43\lesssim\ell\lesssim3.67$.}
At $\ell\approx3.43$, $L_{\rm H}^{\rm same}$ and $L_{\rm branchless}$
in turn cross, giving
\begin{equation}
L_{\rm hybrid}<L_{\rm H}^{\rm same}<L_{\rm branchless}.
\end{equation}
Parameter Choice~2 ($\ell=3.55$) lies within this regime.

\paragraph*{$\ell\gtrsim3.67$.}
Were $\ell$ increased further, one might expect $L_{\rm H}^{\rm
same}$ to eventually overtake $L_{\rm hybrid}$ as well, since the H-tree reoptimizes with $\ell$ on both sides, while the branchless
configuration is linear in $\ell$ and the hybrid reoptimizes on one side only. This is indeed so: we find that $L_{\rm H}^{\rm same}$ does overtake $L_{\rm
hybrid}$, but only at $\ell\approx4.11$. This crossing, however,
comes too late. The bipartite crossing condition of
Eq.~\eqref{eq:crossing-condition}, $2z(\ell,\ell)>2\ell$, is
saturated with equality at the precise root
$\ell_*=3.671274\ldots$, where the bound is realized by an explicit
closed geodesic, the curve $\Gamma_N$ corresponding to the group
element $N=\gamma_2\gamma_2'^{-1}$, which satisfies
$L_{\Gamma_N}(0)=2z$. For $\ell$ slightly beyond $\ell_*$, we have
$L_{\Gamma_N}(0)<2\ell$, so $\Gamma_N$ wins the RT competition for
the partition $AB|A'B'$; by continuity this persists over a range
of small $|\tau|$. Crucially, its length is $\tau$-dependent: $L_{\Gamma_N}(\tau)=2z+c\,\tau^2+O(\tau^4)$ with
$c\approx0.526\neq0$, verified numerically. The winning connected
surface therefore genuinely detects the twist, so the bipartite RT
entropies become twist-sensitive already at $\ell\simeq\ell_*$, well
before $\ell\simeq4.11$.

Thus, within the class of configurations analyzed here, for
sufficiently small $|\tau|$, we find no point along the symmetric
line $\ell=m$ at which a twist-sensitive configuration wins the
global competition while the bipartite RT entropies remain exactly
$\tau$-independent. We emphasize that this specific sequence of
crossings, and their order relative to the bipartite transition, is
not something that could have been anticipated a priori: it is the
outcome of an explicit numerical optimization of the trivalent
vertices at each value of $\ell$, not a consequence derivable from
general principles alone.

\section{VII.  Summary and Discussion}

We have investigated, for a family of four-boundary AdS$_3$
wormholes related by a Fenchel--Nielsen twist $\tau$ along an
internal pants-decomposition curve, whether the complete set of
bipartite RT entropies can remain exactly $\tau$-independent while
the holographic $\mathtt q=4$ multi-entropy detects the twist. At both Parameter Choice~1 ($\ell=m=2$) and Parameter Choice~2 ($\ell=m=3.55$), within the class of admissible configurations
analyzed in this work, the minimal $\mathtt q=4$ network is exactly $\tau$-independent, so $S^{(4)}$ does not detect the twist at either point, and the bipartite RT data are twist-invariant there
as well. At Parameter Choice~1 the minimal network is a branchless configuration; at Parameter Choice~2 it is instead the hybrid configuration of Sec.~V, in which one pair of pants is equipped with a Steiner tripod while the other is cut directly along a horizon.

We emphasize that the twist need not be invisible to other boundary
observables. For example, correlation functions between operators on
different asymptotic boundaries are generically sensitive to the
bulk geodesic distances connecting those boundaries and may therefore
detect the twist.

Within the class of configurations analyzed here, for sufficiently
small $|\tau|$, we have not found a point along the symmetric line
$\ell=m$ at which the holographic $\mathtt q=4$ multi-entropy
detects a bulk modulus invisible to bipartite RT entropies. As
shown in Sec.~VI, this holds all the way to the point $\ell_*\simeq3.671274$ where the bipartite RT entropies
themselves become twist-sensitive, so this conclusion is not an
artifact of an unlucky choice of parameters.

As a brief comment on other partitions, consider the crossing-type
partition $AB|A'|B'$. One candidate consists of the crossing
geodesic $\Gamma_N$ of Sec.~VI together with one horizon, say $A'$.
At $\tau=0$, its length is
\begin{equation}
L_{\Gamma_N}(0)+\ell=2z+\ell.
\end{equation}
A simpler, $\tau$-independent candidate is obtained by cutting the
two horizons $A'$ and $B'$, with total length $2\ell$; the remaining
boundaries $A$ and $B$ then belong to the same region. The first
candidate could beat the second only if
\begin{equation}
2z+\ell<2\ell,
\qquad\Leftrightarrow\qquad
z<\frac{\ell}{2}.
\end{equation}
However, this is impossible in the region where the bipartite RT
entropies are twist-invariant, where $z>\ell$ by
Eq.~\eqref{eq:crossing-condition}. In fact, near
$\tau=0$ the situation only becomes less favorable for the crossing
candidate, since
\begin{equation}
L_{\Gamma_N}(\tau)=2z+c\,\tau^2+O(\tau^4),
\qquad c>0.
\end{equation}
Thus this particular twist-sensitive $\mathtt q=3$ candidate cannot win in
the region of interest, at least near $\tau=0$. We have not attempted
an exhaustive search over $\mathtt q=3$ configurations, so this observation
should be regarded only as a consistency check.

Whether higher-partite entanglement can in fact access bulk information
invisible to bipartite entanglement, in the sense originally motivating
this question \cite{Iizuka:2025bcc}, therefore remains open within the regime analyzed
here; in a companion paper~\cite{Anegawa:2026PartII}, we explore this
question further at finite twist. Our analysis here has been restricted
to a limited region of parameter space, and it would be interesting to
explore more systematically whether some other region of the
Fenchel--Nielsen moduli space, or a multiboundary wormhole with more
asymptotic regions, does exhibit the phenomenon sought here.

A simple heuristic \footnote{We thank T.~Anegawa for pointing out
this observation to us.} makes the competition transparent. All
three lengths share the cuff $M$ (length $m$) once, and differ only
in how each of the two sides, $\{A,A',M\}$ and $\{M,B,B'\}$, is
crudely approximated: either by the $z$-loop (the returning
orthogeodesic of Sec.~III) or by a direct horizon of length $\ell$:
\begin{align}
L_{\rm H}^{\rm heur}
&=\underbrace{m}_{M}
+\underbrace{z}_{\{A,A'\}\text{ side}}
+\underbrace{z}_{\{B,B'\}\text{ side}}
=2z+\ell,
\\
L_{\rm hybrid}^{\rm heur}
&=\underbrace{m}_{M}
+\underbrace{z}_{\{A,A'\}\text{ side}}
+\underbrace{\ell}_{\{B,B'\}\text{ side}}
=z+2\ell,
\\
L_{\rm branchless}
&=\underbrace{m}_{M}
+\underbrace{\ell}_{\{A,A'\}\text{ side}}
+\underbrace{\ell}_{\{B,B'\}\text{ side}}
=3\ell,
\end{align}
where we used $m=\ell$ on the symmetric line. The H-tree connects across the twisted gluing on both sides (hence $z$ on
each), the hybrid replaces only the $\{B,B'\}$ side by the direct
horizon $B$ (Sec.~V), and the branchless configuration replaces
both sides by direct horizons; on the symmetric line $m=\ell$ this coincides with simply cutting three of the four horizons directly. Since each step simply trades one
factor of $z$ for one factor of $\ell$,
\begin{equation}
L_{\rm H}^{\rm heur}-L_{\rm hybrid}^{\rm heur}
=L_{\rm hybrid}^{\rm heur}-L_{\rm branchless}
=z-\ell,
\end{equation}
so that whenever $z>\ell$ (the same condition,
Eq.~\eqref{eq:crossing-condition}, that keeps the bipartite RT
entropies twist-invariant), this heuristic predicts
$L_{\rm branchless}<L_{\rm hybrid}^{\rm heur}<L_{\rm H}^{\rm heur}$
throughout: a naive no-go, in which the only twist-sensitive
configuration is heuristically the longest one everywhere bipartite
RT is blind to the twist. 

However, this naive no-go is not exact: as shown in
Sec.~VI, the actual numerical optimization reveals that $L_{\rm
branchless}$ and $L_{\rm hybrid}$, and separately $L_{\rm
branchless}$ and $L_{\rm H}^{\rm same}$, each cross \emph{within}
the region $z>\ell$ (at $\ell\simeq2.66$ and $\ell\simeq3.43$,
respectively), so the ordering above does not hold exactly. 
The heuristic correctly identifies the mechanism and
the marginal region $z\simeq\ell$, but not the precise ordering
among the three configurations.

\begin{center}
\textbf{Acknowledgments}
\end{center}

\begin{acknowledgments}
We thank T.~Anegawa, A.~Miyata, and K.~Tamaoka for helpful comments and questions on the draft.
The work of N.I. was supported in part 
by  NSTC of Taiwan Grant Number 114-2112-M-007-025-MY3, 
and 
by MEXT KAKENHI Grant-in-Aid for Transformative Research Areas A “Extreme Universe” No. 21H05184. 
The author used Claude (Anthropic, Claude Sonnet 5) to assist with the generation and debugging of numerical optimization code in
Secs.~V--VI, Claude and ChatGPT (OpenAI) to assist with figure preparation, and generative AI tools for language polishing.
All AI-assisted code, numerical results, figures, and text were checked and verified by the author.
\end{acknowledgments}

\vspace{1mm}

\appendix
\begin{center}
\textbf{Appendix}
\end{center}

\vspace{-7mm}

\section{A: Fundamental domain and side-pairing generators}
\label{App:sidepairing}

We verify that $\gamma_2$ pairs $C_1$ and $C_2$ with reversed
orientation. 
Using the matrix representation of $\gamma_2$ given in Eq.~\eqref{gamma2matrixrepn}, the corresponding M\"obius transformation is
\begin{equation}
\gamma_2(z)
=
\frac{-c_2 z+c_1c_2+R_1R_2}{c_1-z}
=
c_2+\frac{R_1R_2}{c_1-z}.
\label{gamma2zformula}
\end{equation}
Parametrizing $C_1$ as $z=c_1+R_1e^{i\theta}$, $\theta\in[0,\pi]$,
gives
\begin{equation}
\gamma_2(z)=c_2+R_2\,e^{i(\pi-\theta)} ,
\end{equation}
so that $\gamma_2$ maps $C_1$ onto $C_2$, but with $\theta\to\pi-\theta$:
as $\theta$ increases from $0$ to $\pi$, the image point traverses
$C_2$ in the opposite sense. Thus $\gamma_2$ pairs $C_1$ with $C_2$ with reversed boundary orientation. The same holds for $\gamma_2':C_1'\to C_2'$.

Quotienting $\mathbb{H}^2$ by $\Gamma=\langle\gamma_1,\gamma_2,\gamma_2'\rangle$
produces the four-boundary surface $\Sigma=\mathbb{H}^2/\Gamma$: 
the identification of $C_a$ and $C_b$ by $\gamma_1$ produces the
internal cuff $M$, while $\gamma_2,\gamma_2'$ pair the circles
nested inside the annulus between $C_a$ and $C_b$. The conjugacy class of each generator
determines a closed geodesic on $\Sigma$: that of $\gamma_2$ is the
horizon $A$, that of $\gamma_2'$ is the horizon $B$, and those of
$\gamma_1\gamma_2$ and $\gamma_1\gamma_2'$ are the horizons $A'$ and
$B'$ respectively.

Since $\gamma_1$ represents the identification along the cuff $M$, 
it is convenient to
choose the fundamental domain so that $\gamma_1$ is diagonal along
its own axis,
\begin{equation}
\gamma_1=
\begin{pmatrix}
e^{L_M/2} & 0\\
0 & e^{-L_M/2}
\end{pmatrix}.
\end{equation}
Cutting $\Sigma$ along $M$ separates it into the two pairs of pants
containing $\{A,A'\}$ and $\{B,B'\}$ respectively. 

The Fenchel--Nielsen twist described in the main text,
$\gamma_2'\to\eta(\tau)\gamma_2'\eta(-\tau)$, is precisely the
operation of translating the second pair of pants along the axis of
$\gamma_1$ by signed length $\tau$ before regluing along $M$.

Explicitly, since $\eta(\tau)$ acts on $\mathbb{H}^2$ as
$z\mapsto e^\tau z$, the conjugated generator $\eta(\tau)\gamma_2'\eta(-\tau)$
pairs the uniformly rescaled circles, with
\[
(c_1',c_2',R_1',R_2')
\longrightarrow
(e^\tau c_1',e^\tau c_2',e^\tau R_1',e^\tau R_2'),
\]
namely,
\begin{equation}
\gamma_2'(\tau)(z)=e^\tau c_2'+\frac{e^{2\tau}R_1'R_2'}{e^\tau c_1'-z}.
\end{equation}
This rescaling in the upper-half-plane coordinates should not be
confused with a change of the intrinsic geometry of the second pair
of pants: $z\mapsto e^\tau z$ is a hyperbolic isometry, corresponding
to a translation by signed length $\tau$ along the axis of
$\gamma_1$. Thus the intrinsic boundary lengths remain unchanged;
only the relative gluing along $M$ is twisted.

\section{B: Trace and geodesic length}\label{App:tracelength}

Here we derive the formula~\eqref{tracelengthformula}, relating the
trace of a hyperbolic element $\gamma\in SL(2,\mathbb{R})$ to the
length of its associated closed geodesic. Since $\gamma$ and $-\gamma$ induce the same M\"obius action on $z$
(the isometry group of $\mathbb H^2$ being $PSL(2,\mathbb
R)=SL(2,\mathbb R)/\{\pm I\}$), the geodesic length can depend on
$\gamma$ only through the sign-independent combination $|\Tr\gamma|$. A hyperbolic
element thus satisfies $|\Tr\gamma|>2$. Writing
$\gamma=\begin{pmatrix}a&b\\c&d\end{pmatrix}$ with $ad-bc=1$, its
fixed points on $\partial\mathbb{H}^2=\mathbb{R}\cup\{\infty\}$
satisfy
\begin{equation}
cz^2+(d-a)z-b=0,
\end{equation}
whose discriminant is
\begin{equation}
\Delta=(d-a)^2+4bc=(\Tr\gamma)^2-4>0.
\end{equation}
Hence $\gamma$ has two distinct real fixed points $x_-,x_+$, and the
geodesic joining them, the semicircle orthogonal to the real axis
(or a vertical line if one endpoint is at infinity), is the
invariant axis of $\gamma$.

To put the invariant axis in a standard form, we map its two
endpoints to $0$ and $\infty$. Choose
$h\in PSL(2,\mathbb{R})$ with
$h(x_-)=0$ and $h(x_+)=\infty$. For example, one may take
\[
h(z)=\frac{z-x_-}{x_+-z},
\]
up to an overall normalization of the representing matrix, so that
\(h\in PSL(2,\mathbb R)\).
The conjugate element
$\widetilde\gamma=h\gamma h^{-1}$ fixes both $0$ and $\infty$.
It must therefore be diagonal,
\begin{equation}
\widetilde\gamma=
\begin{pmatrix}
\lambda&0\\0&\lambda^{-1}
\end{pmatrix},
\qquad
\lambda>1.
\end{equation}
This $\widetilde\gamma$ acts as $z\mapsto\lambda^2 z$, with invariant axis the imaginary
axis. Along this axis, $ds=dy/y$, so the hyperbolic distance
traveled between $iy$ and $i\lambda^2y$ is
\begin{equation}
L(\gamma)
=
\int_{y}^{\lambda^2y}\frac{dy'}{y'}
=
2\log\lambda .
\end{equation}
Since conjugation by an isometry preserves translation length, this
is also the translation length of the original $\gamma$ along its
own invariant axis. Trace is likewise invariant under conjugation,
so
\begin{equation}
|\Tr\gamma|
=
|\Tr\widetilde\gamma|
=
\lambda+\lambda^{-1}
=
2\cosh\frac{L(\gamma)}{2},
\end{equation}
which is the formula used in Eq.~\eqref{tracelengthformula}.
See also Ref.~\cite{Maxfield:2014kra}.

\section{C: Distance between two disjoint geodesics in $\mathbb H^2$}\label{App:distance}

Here we derive the formula used in the main text for the distance
between the two disjoint geodesics with ideal endpoints
$(0,\infty)$ and $(r,s)$, where
\[
0<r<s .
\]

We work in the upper half-plane metric
\begin{equation}
ds^2=\frac{dx^2+dy^2}{y^2}.
\end{equation}

The geodesic with endpoints $(0,\infty)$ is the imaginary axis,
while the geodesic with endpoints $(r,s)$ is the semicircle
\begin{equation}
\left(x-\frac{r+s}{2}\right)^2+y^2
=
\left(\frac{s-r}{2}\right)^2 .
\end{equation}

A geodesic perpendicular to the imaginary axis must be a semicircle centered at the origin. Furthermore, let its Euclidean radius be $\rho$. Then, orthogonality
to the semicircle $(r,s)$ requires
\begin{equation}
\rho^2 + \left(\frac{s-r}{2}\right)^2 = \left(\frac{r+s}{2}\right)^2,
\end{equation}
and hence
\begin{equation}
\rho=\sqrt{rs}.
\end{equation}

We parameterize the semicircle centered at the origin by
\begin{equation}
x=\rho\cos\theta,
\qquad
y=\rho\sin\theta .
\end{equation}
Along this curve,
\begin{equation}
ds
=
\frac{\sqrt{dx^2+dy^2}}{y}
=
\frac{d\theta}{\sin\theta}.
\end{equation}

Its intersection with the imaginary axis occurs at
$\theta=\pi/2$. Let $\theta_*$ denote its intersection with the
semicircle geodesic $(r,s)$. Using the two circle equations, one finds
\begin{equation}
\cos\theta_*
=
\frac{2\sqrt{rs}}{r+s},
\qquad
\sin\theta_*
=
\frac{s-r}{r+s}.
\end{equation}
Therefore the hyperbolic length of the common perpendicular is
\begin{equation}
d
=
\int_{\theta_*}^{\pi/2}\frac{d\theta}{\sin\theta}
=
\operatorname{arctanh}(\cos\theta_*)
=
\operatorname{arctanh}\!\left(
\frac{2\sqrt{rs}}{r+s}
\right).
\end{equation}
Using
\begin{equation}
\operatorname{arctanh} x
=
\frac12\log\!\left(\frac{1+x}{1-x}\right), \qquad |x|<1,
\end{equation}
we obtain
\begin{equation}
d
=
\frac12
\log\!\left(
\frac{r+s+2\sqrt{rs}}
{r+s-2\sqrt{rs}}
\right)
=
\log\!\left(
\frac{\sqrt{s}+\sqrt{r}}
{\sqrt{s}-\sqrt{r}}
\right),
\end{equation}
where we have used \(0<r<s\).

Equivalently,
\begin{equation}
e^d=\frac{\sqrt{s}+\sqrt{r}}{\sqrt{s}-\sqrt{r}}, 
\end{equation}
or,  
\begin{equation}
\cosh d
=
\frac{r+s}{s-r},
\end{equation}
thus, we obtain 
\begin{equation}
d
=
\operatorname{arcosh}\!\left(
\frac{r+s}{s-r}
\right).
\end{equation}

\bibliographystyle{apsrev4-2}
\nocite{Balasubramanian:2024ysu}
\bibliography{reference}

\end{document}